\documentclass[twocolumn,epjc3]{svjour3}          
\usepackage{cuted}
\usepackage{amsmath}
\RequirePackage[T1]{fontenc}

\smartqed  

\RequirePackage{graphicx}
\RequirePackage{mathptmx}      
\RequirePackage{flushend}
\RequirePackage[numbers,sort&compress]{natbib}
\RequirePackage[colorlinks,citecolor=blue,urlcolor=blue,linkcolor=blue]{hyperref}

\journalname{Eur. Phys. J. C}

\begin{document}

\title{Detectability of Gravitational Waves from Extreme-Mass-Ratio Inspirals on Periodic Orbits around Bardeen Black Holes}


\author{Li Huang\thanksref{e1,addr1}
        \and
        Xing He\thanksref{e2,addr2} 
}

\thankstext{e1}{e-mail: lihuang@ynnu.edu.cn}
\thankstext{e2}{e-mail: xinghe76327@163.com}

\institute{School of Physics and Electronic Information, Yunnan Normal University, Kunming 650500, People's Republic China\label{addr1}
          \and
          School of Mathematics and Physics, Shangrao Normal University, Shangrao 334001, People's Republic China\label{addr2}
}

\date{Received: date / Accepted: date}

\maketitle

\begin{abstract}
We investigate gravitational wave (GW) signatures from extreme-mass-ratio inspirals (EMRIs) around a Bardeen black hole, modeling the small object as a test particle, and provide a framework for probing strong-field spacetime geometry. We first present the equations of motion for test particles around a Bardeen black hole and characterize circular orbits in the parameter space of energy and angular momentum. Using Levin's classification, we adopt the extended classification scheme to classify periodic orbits around a Bardeen black hole. We then compute GW waveforms in both the time and frequency domains, as well as the corresponding characteristic strains, for these periodic orbits and compare them with the sensitivity curves of the Laser Interferometer Space Antenna (LISA), the evolved Laser Interferometer Space Antenna (eLISA), the TianQin mission, the Big Bang Observer (BBO), and the Decihertz Interferometer Gravitational wave Observatory (DECIGO). We find that detectability is controlled primarily by the luminosity distance. At a luminosity distance of 200 Mpc, the signals can exceed the sensitivity curves of BBO and DECIGO and are comparable to the LISA sensitivity curve, whereas at 2 Gpc, they remain below the sensitivity curves of all detectors considered. These results suggest that EMRIs on periodic orbits around Bardeen black holes offer a promising avenue to probe strong-field gravity and the properties of Bardeen black holes.
\end{abstract}

\section{Introduction} 
\label{sec:intro}
Einstein's general relativity (GR) and quantum mechanics constitute two fundamental pillars of modern physics~\cite{Einstein1916AnP354.769,Dirac1928RSPSA117.610}. 
To date, GR has successfully passed numerous precision tests, confirming its robustness in describing gravitational phenomena~\cite{Will2014LRR17.4} across a wide range of scales, from cosmological observations by the Planck collaboration~\cite{Planck2020aap641.A1} to the dynamics of binary pulsars~\cite{Hulse1975ApJ195.L51,Weisberg1981GRG13.1}. However, unresolved issues such as spacetime singularities and the absence of a complete theory of quantum gravity have motivated various modifications to the theory~\cite{Bardeen1968qtr87,Rovelli1998LRR1.1}.

These proposed extensions often aim to resolve the inconsistencies between GR and quantum mechanics, particularly at small scales or high energies where a unified framework is still lacking. A comprehensive review of regular black hole models and their observational implications can be found in~\cite{Bambi2023arXiv230713249}. Among these proposals, the Bardeen black hole spacetime stands out as a particularly well-motivated candidate for a regular black hole. Originally introduced as a regular black hole solution coupled to nonlinear electrodynamics, it avoids the central singularity while preserving a horizon structure that closely mimics the Schwarzschild black hole in the weak-field limit, with deviations controlled by a magnetic monopole charge~\cite{Bardeen1968qtr87,Stuchlik2015IJMPD2450020}.
Its geodesic structure and shadow properties have been extensively studied~\cite{Stuchlik2015IJMPD2450020,Bambi2023arXiv230713249}, making it an ideal testing ground for strong-field gravity and possible deviations from GR.

Until recently, direct tests of strong-field gravity around compact objects remained elusive. This changed with the groundbreaking detection of gravitational waves (GWs) from a binary black hole merger~\cite{Abbott2016PRL116.061102}, opening a new frontier for testing GR in extreme environments. With the LIGO-Virgo-KAGRA (LVK) network of ground-based interferometers~\cite{Abbott2023PRX13.041039,2021arXiv211206861T} and future ground-based observatories such as the Einstein Telescope (ET)~\cite{Colpi2024arXiv2402.07571,Branchesi2023JCAP07.068}, as well as planned space-based detectors including LISA~\cite{Robson2019CQG36.105011,Barausse2020GRG52.81,Auclair2023LRR26.5,Arun2022LRR25.4,Colpi2024arXiv2402.07571}, TianQin~\cite{Luo2016CQG33.035010}, Taiji~\cite{Huang2017SSPMA47.010404,Du2026SCPMA69.249501}, Big Bang Observer (BBO)~\cite{Cutler2006PRD73.042001}, and Decihertz Interferometer Gravitational wave Observatory (DECIGO)~\cite{Seto2001PRL87.221103,DECIGO2021PTEP2021.05A105}, we are now able to probe GR in unprecedented detail. These advanced instruments will enable tests of GR under extreme conditions and explorations of possible modifications, including extensions toward quantum gravity~\cite{Mirshekari2012PRD85.024041,Canizares2012PRD86.044010,Ellis2016MPLA31.1675001,Yunes2016PRD94.084002,Arzano2016PRD93.l24065,Calcagni2019PLB798.135000,Arun2022LRR25.4,Addazi2022PPNP125.103948,Calcagni2022arXiv2206.07066}. Among the most promising GW sources for testing strong-field gravity are extreme-mass-ratio inspirals (EMRIs), in which a stellar-mass compact object spirals into a massive black hole. During its long inspiral, the small body traces out a sequence of nearly periodic orbits, making the orbital dynamics and the resulting gravitational radiation highly sensitive to the underlying spacetime geometry~\cite{Amaro-Seoane2017arXiv1702.00786,Barack2019CQG36.143001}. This sensitivity turns EMRIs into precision probes of black hole spacetimes and potential deviations from GR in the strong-field regime.

For this reason, a substantial part of this paper is devoted to the geodesic structure and associated dynamics of black hole spacetimes, both within classical GR and in potential extensions thereof. Understanding the motion of test particles and the propagation of light in strong-field backgrounds provides essential theoretical predictions that can be compared directly with current and future observational data. The investigation of geodesic structures reveals intrinsic properties of black hole spacetimes—such as horizon structure, innermost stable circular orbits (ISCOs), and photon spheres—and informs us about possible deviations from GR predicted by alternative theories of gravity. For Bardeen black hole spacetimes, the circular geodesic structure has been analyzed in detail in~\cite{Stuchlik2015IJMPD2450020}, and bound orbits between the ISCO and the marginally bound orbit (MBO) have been systematically studied using a periodic orbit classification~\cite{Stuchlik2015IJMPD2450020,Gao2020AP418.168194}. While these works have established the orbital classification and stability properties of bound geodesics in Bardeen black hole spacetimes, the GW signatures of such periodic orbits and their detectability with space-based detectors remain largely unexplored. The present paper aims to fill this gap by computing the characteristic strains of GWs emitted by periodic orbits around Bardeen black holes and comparing them directly with the sensitivity curves of planned space-based detectors.

To further characterize particle motion in strong-field spacetimes, we adopt the extended periodic orbit classification $(z,w,v;e)$ introduced in~\cite{Lim2024PRD109.024037,Chan2025GRG57.35,Huang2025PRD111.084038}. Periodic orbits provide a systematic probe of the underlying spacetime geometry, allowing us to analyze orbital stability, resonances, and their imprint on the emitted gravitational radiation~\cite{Huang2025JCAP10.067,Li2025arXiv250907333}. Recent studies have applied this framework to various regular black hole spacetimes, demonstrating that the characteristic strains from periodic orbits carry distinctive signatures of the underlying geometry and can exceed the sensitivity curves of space-based detectors such as LISA in the millihertz band~\cite{Hua2026PLB876.140402,Gong2026EPJC86.469}.
By computing the characteristic strains of these periodic orbit GW signals and comparing them with the sensitivity curves of planned space-based detectors~\cite{Blanchet2024LRR27.4,Abbott2016PRL116.061102,Luo2016CQG33.035010,Abbott2019PRX9.031040}, we assess their detectability and their potential for constraining the properties of black holes.

The paper is organized as follows. In Sec.~\ref{sec:2}, we derive the geodesic equations using the Lagrangian formalism and characterize circular orbits in the $(L,E)$ parameter space for the Bardeen black-hole spacetime. In Sec.~\ref{sec:3}, we present an analytic parametrization of periodic orbits in terms of the orbital eccentricity $e$ and latus rectum $\lambda$, derive the periodicity condition, and construct periodic orbits with prescribed $(z,w,v;e)$. In Sec.~\ref{sec:4}, we compute the GW waveforms and
characteristic strains in both the time and frequency domains and compare them with the sensitivity curves of space-based detectors. In Sec.~\ref{sec:5}, we summarize our results and discuss possible directions for future work. We
adopt the metric signature $(-,+,+,+)$ and use geometrized units with $G=c=\hbar=1$.

\section{Equations of Motion and the $(L,E)$ Parameter Space}
\label{sec:2}

In this section, we revisit the derivation of geodesics for timelike test particles using the Lagrangian formalism. We derive analytical expressions for the conserved energy $E$ and angular momentum $L$ in terms of the circular orbit radius and the orbital parameters $e$ and $\lambda$. These expressions are used to characterize periodic orbits, particularly through their representation in the $(L,E)$ parameter space.

\subsection{Geodesic Equations from the Lagrangian Formalism}
\label{sec:2.1}

The Bardeen black hole is a static, spherically symmetric spacetime described by the metric line element
\begin{align}
\label{metric}
\mathrm{d}s^2
&=g_{\mu\nu}\mathrm{d}x^{\mu}\mathrm{d}x^{\nu}
\nonumber \\
&=-A(r)\mathrm{d}t^2+\frac{1}{A(r)}\mathrm{d}r^2
+B(r)\left(\mathrm{d}\theta^2+\sin^2\theta\,\mathrm{d}\phi^2\right),
\nonumber \\
\end{align}
with metric coefficients
\begin{align}
\label{aarr}
A(r)&=1-\frac{2Mr^2}{(r^2+g^2)^{3/2}},\\
\label{ccrr}
B(r)&=r^2.
\end{align}
Here, $M$ denotes the black hole mass, while the parameter $g$ is related to the magnetic monopole charge in nonlinear electrodynamics. This charge leads to phenomena similar to those found in charged black hole spacetimes.
Equation~(\ref{metric}) describes a singularity-free black hole satisfying the condition~\cite{Ayonbeato2000PLB493.149}
\begin{align}
\label{npf}
g^2 \leq \frac{16}{27}M^2.
\end{align}
The analysis in Ref.~\cite{Gao2020AP418.168194} indicates the following horizon structure:
\begin{enumerate}
\item two horizons when $g^2<\frac{16}{27}M^2$;
\item one horizon when $g^2=\frac{16}{27}M^2$;
\item no horizons when $g^2>\frac{16}{27}M^2$.
\end{enumerate}
It is worth emphasizing that, for $g=0$, the spacetime reduces to the Schwarzschild solution. For a timelike test particle around a Bardeen black hole spacetime, the Lagrangian governing its geodesic motion is
\begin{align}
\label{LLLL}
\mathcal{L}(x,\dot{x})=\frac{1}{2}g_{\mu\nu}\dot{x}^{\mu}\dot{x}^{\nu},
\end{align}
where $x^{\mu}=(t,r,\theta,\phi)$ denotes the coordinates of the particle along its timelike worldline, and an overdot denotes differentiation with respect to the proper time $\tau$. In the Bardeen black hole spacetime, this Lagrangian takes the explicit form
\begin{align}
\label{Lagrange}
\mathcal{L}(x,\dot{x})=\frac{1}{2}\left(-A(r)\dot{t}^{2}
+\frac{\dot{r}^{2}}{A(r)}+r^{2}\dot{\theta}^{2}
+r^{2}\sin^{2}\theta\,\dot{\phi}^{2}\right),
\end{align}
where $(\dot{t},\dot{r},\dot{\theta},\dot{\phi})$ are the components of the particle's four-velocity. This Lagrangian describes the motion of a test particle in the gravitational field of the Bardeen black hole, which is a regular, singularity-free alternative to the Schwarz\-schild solution.

The canonical momenta are defined by $p_{\mu}=\partial \mathcal{L}/\partial\dot{x}^{\mu}$, which gives
\begin{align}
\label{pptt}
p_t&=-A(r)\dot{t},                   
\\
\label{pprr}
p_r&=\frac{\dot{r}}{A(r)},           
\\
\label{ptheta}
p_{\theta}&=r^2\dot{\theta},         
\\
\label{pphi}
p_{\phi}&=r^2\sin^2\theta\,\dot{\phi}.  
\end{align}
The corresponding Hamiltonian can be derived by a Legendre transformation of the Lagrangian. The \mbox{Hamilton-Jacobi} equation for the system is given by
\begin{align}
\label{hjeq}
&\frac{1}{2}\Bigg[-\frac{1}{A(r)}\left(\frac{\partial S}{\partial t}\right)^2+A(r)\left(\frac{\partial S}{\partial r}\right)^2+\frac{1}{r^2}\left(\frac{\partial S}{\partial\theta}\right)^2
\nonumber    \\
&+\frac{1}{r^2\sin^2\theta}\left(\frac{\partial S}{\partial\phi}\right)^2\Bigg]+\frac{\partial S}{\partial \tau}=0.  
\end{align}
For timelike test particles around a Bardeen black hole, we consider, without loss of generality, motion confined to the equatorial plane, $\theta=\pi/2$. The static and spherically symmetric nature of the spacetime gives rise to two conserved quantities, namely the energy $E$ and angular momentum $L$, which are given by
\begin{align}
\label{dtthta}
\dot{t}&=\frac{E}{A(r)},
\\
\dot{\phi}&=\frac{L}{r^2}.
\end{align}
The radial equation of motion is then
\begin{align}
\label{radione}
\dot{r}^2=E^2-V_{\mathrm{eff}}(r),
\end{align}
where $V_{\mathrm{eff}}(r)$ denotes the effective potential governing the radial motion, given by
\begin{align}
\label{veff}
V_{\mathrm{eff}}=\left(1+\frac{L^2}{r^2}\right)
\left[1-\frac{2Mr^2}{(r^2+g^2)^{3/2}}\right].  
\end{align}
The radial acceleration $\ddot{r}$ can then be obtained from the radial equation of motion as
\begin{align}
\label{raditwo}
\ddot{r}=\frac{A'(r)\dot{r}^2}{2A(r)}-\frac{A'(r)E^2}{2A(r)}+\frac{L^2A(r)}{r^3}.  
\end{align}
To obtain the orbit equation relating $r$ and $\phi$, we use $\dot{r}/\dot{\phi}=\mathrm{d}r/\mathrm{d}\phi$, which yields
\begin{align}
\label{drdphi}
\frac{\mathrm{d}r}{\mathrm{d}\phi}=\pm r^2\sqrt{\frac{E^2-1}{L^2}+\frac{2Mr^2}{L^2(r^2+g^2)^{3/2}}-\frac{1}{r^2}+\frac{2M}{(r^2+g^2)^{3/2}}}.  
\nonumber  \\
\end{align}

\subsection{Characterizing Circular Orbits in the $(L,E)$ Parameter Space}
\label{sec:2.2}

In the classification of periodic orbits~\cite{Levin2008PRD77.103005,Levin2009PRD79.043016,Lim2024PRD109.024037,Chan2025GRG57.35}, circular orbits are regarded as the zero eccentricity limit. For a timelike test particle around a Bardeen black hole, the angular momentum $L$ and energy $E$ can be expressed as functions of the circular orbit radius $r_\mathrm{c}$. The conditions for circular orbits are
\begin{align}
\label{ygd}
E^2&=V_{\mathrm{eff}},   
\nonumber \\
V'_{\mathrm{eff}}&=0. 
\end{align}
Solving these equations for $E$ and $L$, we obtain
\begin{align}
\label{enzhi}
E(r_\mathrm{c},g)&=\sqrt{\frac{(r_\mathrm{c}^2+g^2)^{5/2}-2Mr_\mathrm{c}^4-2Mr_\mathrm{c}^2g^2}{(r_\mathrm{c}^2+g^2)^{5/2}-3Mr_\mathrm{c}^4}}
\nonumber  \\
&\quad\times\sqrt{\frac{(r_\mathrm{c}^2+g^2)^{3/2}-2Mr_\mathrm{c}^2}{(r_\mathrm{c}^2+g^2)^{3/2}}}, 
\\
\label{jiaozhi}
L(r_\mathrm{c},g)&=r_\mathrm{c}\sqrt{\frac{Mr_\mathrm{c}^4-2Mr_\mathrm{c}^2g^2}{(r_\mathrm{c}^2+g^2)^{5/2}-3Mr_\mathrm{c}^4}}.
\end{align}
For $g=0$, Eqs.~(\ref{enzhi}) and~(\ref{jiaozhi}) reduce, respectively, to
\begin{align}
\label{enzhitwo}
E(r_\mathrm{c},0)&=\frac{r_\mathrm{c}-2M}{\sqrt{r_\mathrm{c}(r_\mathrm{c}-3M)}}, 
\\
\label{jiaozhitwo}
L(r_\mathrm{c},0)&=r_\mathrm{c}\sqrt{\frac{M}{r_\mathrm{c}-3M}},
\end{align}
in agreement with Ref.~\cite{Lim2024PRD109.024037}.
This is the Schwarzschild case, In this case, the Bardeen black hole reduces to the Schwarzschild black hole.

To determine the stability of circular orbits, we evaluate the second derivative of the effective potential $V_{\mathrm{eff}}$ at $r=r_\mathrm{c}$, which yields

\begin{align}
\label{veffer}
V''_{\mathrm{eff}}(r_\mathrm{c},g)&=\frac{18Mr_\mathrm{c}^2(r_\mathrm{c}^2+g^2)-30Mr_\mathrm{c}^4}{(r_\mathrm{c}^2+g^2)^{7/2}}
\nonumber \\
&\quad-\frac{4M(r_\mathrm{c}^2+g^2)-12Mr_\mathrm{c}^2}{(r_\mathrm{c}^2+g^2)^{5/2}}
\nonumber \\
&\quad+\frac{6r_\mathrm{c}^2M(r_\mathrm{c}^4-2r_\mathrm{c}^2g^2)}{r_\mathrm{c}^4(r_\mathrm{c}^2+g^2)^{5/2}-3Mr_\mathrm{c}^8}
\nonumber \\
&\quad+\frac{6M^2r_\mathrm{c}^2(g^2-4r_\mathrm{c}^2)(r_\mathrm{c}^4-2r_\mathrm{c}^4g^2)}{(r_\mathrm{c}^2+g^2)^6-3Mr_\mathrm{c}^4(r_\mathrm{c}^2+g^2)^{7/2}}.
\end{align}
Therefore, a circular orbit is stable if
$V''_{\mathrm{eff}}(r_\mathrm{c},g)>0$. It is unstable if
$V''_{\mathrm{eff}}(r_\mathrm{c},g)<0$. The critical points are determined by
\begin{align}
\label{verling}
V''_{\mathrm{eff}}(r_\mathrm{c},g)=0,
\end{align}
at which stability changes between the stable and unstable branches of circular orbits. In particular, the ISCO marks the inner boundary of the stable branch of circular orbits.
This stability criterion is essential for characterizing the orbital motion of timelike test particles around a Bardeen black hole.

The orbital eccentricity $e$ distinguishes between bound and unbound orbits. Bound orbits satisfy $0\leq e<1$. Here, $e=0$ corresponds to circular orbits, and $0<e<1$ corresponds to elliptical orbits. The marginal case $e=1$ corresponds to a parabolic orbit. Unbound orbits with $e>1$ follow hyperbolic trajectories. The stability of the circular orbits along each branch of the parametric curve is determined by the sign of the second derivative of the effective potential, $V''_{\mathrm{eff}}(r_{\mathrm{c}},g)$. We obtain the corresponding values of $(L,E)$ by substituting different values of $r_{\mathrm{c}}$ into Eqs.~(\ref{enzhi})--(\ref{jiaozhi}) for various values of $g$.

\begin{figure}
\includegraphics[width=0.50\textwidth]{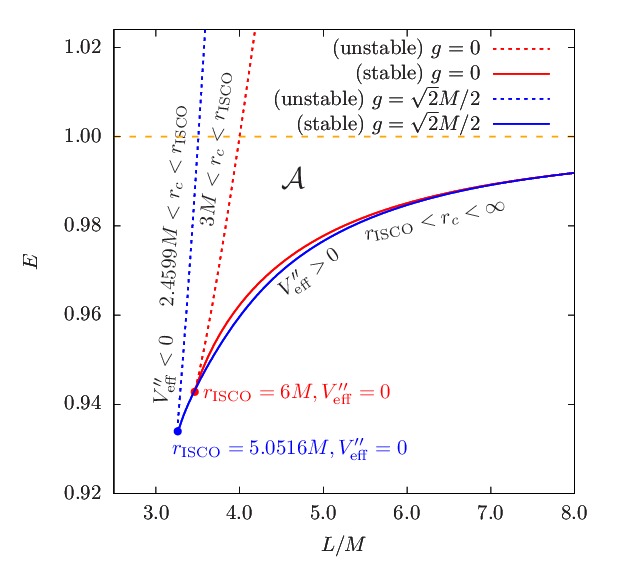}
\caption{The $(L,E)$ parameter space of circular orbits of timelike test particles around a Bardeen black hole for different values of $g$. The domain $\mathcal{A}$ is bounded by the two branches of circular orbits and the line $E=1$. The orange dashed line separates the energetically bound orbit region, $E<1$, from the unbound orbit region, $E>1$.
\label{fig:1}}
\end{figure}

Figure~\ref{fig:1} shows the parametric curves of circular orbits of timelike test particles around Bardeen black holes in the $(L,E)$ plane. The corresponding curve for a Schwarzschild black hole is also shown ($g=0$). The dashed lines (red and green) represent the unstable branches of circular orbits, while the solid lines (red and green) represent the stable branches. Let $\mathcal{A}$ denote the set of $(L,E)$ values associated with bound orbits. In the $(L,E)$ plane, $\mathcal{A}$ is bounded by the circular orbit curves and the line $E=1$, which is shown as an orange dashed curve. Within this region, the equation $E^2=V_{\mathrm{eff}}(r)$ has three distinct real roots. Since the radial motion is allowed where $E^2\geq V_{\mathrm{eff}}(r)$, the particle can oscillate between two adjacent radial turning points. Thus, it remains confined to a finite radial interval and does not cross the event horizon. This defines the region of bound orbits in the $(L,E)$ parameter space.

A bound orbit exhibits periodic radial motion. However, its periapsis generally precesses from one radial cycle to the next. In the presence of periapsis precession, the orbit is generally not a closed ellipse; nevertheless, a closed orbit can occur when the ratio of the radial and azimuthal frequencies is rational. Since periodic motion requires a finite radial range, all periodic orbits considered here are bound. Their corresponding values of $(L,E)$ lie within $\mathcal{A}$. The nonrelativistic limit is recovered as $r_{\mathrm{c}}\rightarrow\infty$. For $g=0$, the Bardeen black hole reduces to the Schwarzschild black hole. 
As shown in Fig.~\ref{fig:1}, the circular orbits are stable, with $V''_{\mathrm{eff}}>0$, for $r_{\mathrm{c}}>6M$.
They are unstable, with $V''_{\mathrm{eff}}<0$, for $3M<r_{\mathrm{c}}<6M$. The stability transition occurs at $r_{\mathrm{c}}=6M$, which corresponds to the ISCO of the Schwarzschild black hole. For $g=\sqrt{2}M/2$, the ISCO is located at $r_{\mathrm{c}}=5.0516M$. Circular orbits with
$r_{\mathrm{c}}>5.0516M$ satisfy $V''_{\mathrm{eff}}>0$ and are therefore stable. Those with $2.4599M<r_{\mathrm{c}}<5.0516M$ satisfy $V''_{\mathrm{eff}}<0$ and are unstable. Compared with the $g=0$ case, the region $\mathcal{A}$ occupies a larger domain in the $(L,E)$ plane.

\section{Distribution of Periodic Orbits in the $(L,E)$ Parameter Space}
\label{sec:3}

Levin and Perez-Giz classified periodic orbits in Refs.~\cite{Levin2008PRD77.103005,Levin2009PRD79.043016}. In this work, we provide a method for determining the angular momentum $L$ and energy $E$ of an orbit specified by $(z,w,v)$. We first briefly describe the procedure for constructing periodic orbits, then derive the expression for the azimuthal angle in terms of the orbital eccentricity, and finally show how to construct a periodic orbit with prescribed values of $(z,w,v)$ and $e$.

Introducing $r=1/u$ and substituting this relation into
Eq.~(\ref{drdphi}) yields
\begin{align}
\label{drduone}
\frac{\mathrm{d}u}{\mathrm{d}\phi}=\pm\sqrt{Q(u)}, 
\end{align}
and
\begin{align}
\label{quone}
Q(u)&=-\frac{1-E^2}{L^2}+\frac{2M}{L^2}\frac{1}{u^2\left(\frac{1}{u^2}+g^2\right)^{3/2}}-u^2
\nonumber   \\
&\quad+2M\frac{1}{\left(\frac{1}{u^2}+g^2\right)^{3/2}},
\end{align}
where $Q(u)$ is an algebraic function of $u$ because of the factor $(1+g^2u^2)^{-3/2}$. Clearly, $u$ is physically admissible only if $Q(u)\geq0$. The values of $u$ satisfying $Q(u)=0$ determine the boundaries of the physically allowed domains.

\subsection{Analytical Parametrization by the Orbital Eccentricity and Latus Rectum} 
\label{sec:3.1}

To fully understand the periodic orbit diagram in $(L,E)$ parameter space, we must also express it in terms of $e$ and $\lambda$. This can be achieved by first re-parameterizing the $u$-root using these parameters.
Based on Eqs.~(\ref{drduone}) and~(\ref{quone}), we have
\begin{equation}
\label{drdutt}
\begin{split}
\frac{\mathrm{d}u}{\mathrm{d}\phi} &= \pm\sqrt{Q(u)} \\
&= \pm\Bigg(\frac{E^2-1}{L^2}+\frac{2M}{L^2}\frac{u}{\left(1+g^2u^2\right)^{3/2}}-u^2 \\
&\qquad+2M\frac{u^3}{\left(1+g^2u^2\right)^{3/2}} \Bigg)^{1/2}.
\end{split}
\end{equation}
As a consistency check, for $g=0$, where the Bardeen black hole reduces to the Schwarzschild black hole, Eq.~(\ref{drdutt}) becomes
\begin{align}
\label{drduthree}
\frac{\mathrm{d}u}{\mathrm{d}\phi}&=\pm\sqrt{Q(u)}
\nonumber   \\
&=\pm\sqrt{\frac{E^2-1}{L^2}+\frac{2M}{L^2}u-u^2+2Mu^3},
\end{align}
This result is consistent with that obtained in Ref.~\cite{Lim2024PRD109.024037}.

For the Bardeen black hole, the roots of the equation $Q(u)=0$ represent the turning points of $r$ in the radial equation of motion. The algebraic function $Q(u)$ can also be rewritten as
\begin{align}
\label{drdufour}
Q(u)=\frac{2M}{\left(1+g^2u^2\right)^{3/2}}(u-u_1)(u_2-u)(u_3-u),
\end{align}
where $u_1$, $u_2$ and $u_3$ are the three relevant real roots. An additional root arising from the algebraic equation obtained after clearing the denominator is discarded because it lies outside the physically admissible domain. Accordingly,
\begin{align}
\label{drdufive}
\frac{\mathrm{d}u}
{\mathrm{d}\phi}&=\pm\sqrt{Q(u)}
\nonumber   \\
&=\pm\sqrt{\frac{2M}{\left(1+g^2u^2\right)^{3/2}}(u-u_1)(u_2-u)(u_3-u)}.
\end{align}
We set the initial condition $\phi=0$ at $u=u_1$. To ensure $L>0$, we take the positive root of Eq.~(\ref{drdufive}), so that $u$ moves from $u_1$ to $u_2$ with $\mathrm{d}u/\mathrm{d}\phi>0$. Integrating Eq.~(\ref{drdufive}) then yields

\begin{align}
\label{drdusix}
\phi=\int_{u_1}^{u}\sqrt{\frac{\left(1+g^2u'^2\right)^{3/2}}{2M}}\frac{\mathrm{d}u'}{\sqrt{(u'-u_1)(u_2-u')(u_3-u')}},
\nonumber \\
\end{align}
In general, the factor $(1+g^2u'^2)^{-3/2}$ prevents the integral in Eq.~(\ref{drdusix}) from being reduced directly to a standard elliptic integral. To proceed analytically, we follow the approach adopted in Refs.~\cite{Chandrasekhar1992book,Lim2024PRD109.024037,Huang2025PRD111.084038} and treat the geometric factor as being evaluated at the turning point $u=u_2$ when extracting the period. Under this approximation, Eq.~(\ref{drdusix}) gives

\begin{align}
\label{drduqi}
\phi(u)&\simeq\frac{2}{\sqrt{\frac{2M}{(1+g^2u_2^2)^{3/2}}(u_3-u_1)}}
\nonumber \\
&\quad\times F\left(\arcsin\sqrt{\frac{u-u_1}{u_2-u_1}},\sqrt{\frac{u_2-u_1}{u_3-u_1}}\right),
\end{align}
where $F(x,k)$ is the incomplete elliptic integral of the first kind. A more accurate treatment can be obtained by evaluating the integral numerically, the analytical expressions below should be understood in the same approximation.

According to Refs.~\cite{Chandrasekhar1992book,Lim2024PRD109.024037,Huang2025PRD111.084038} and the root formula for a cubic equation, the roots are given by
\begin{align}
\label{rootone}
u_1&=\frac{1-e}{\lambda},
\\
\label{roottwo}
u_2&=\frac{1+e}{\lambda},
\\
\label{rootthree}
u_3&=\frac{\left(1+g^2u_3^2\right)^{3/2}}{2M}-\frac{2}{\lambda},
\end{align}
where $\lambda$ is the latus rectum and $e$ is the orbital eccentricity. It is worth emphasizing that Eq.~(\ref{rootthree}) is the equation satisfied by $u_3$, not an explicit solution. Here, we adopt the definition of $e$
from Ref.~\cite{Levin2008PRD77.103005}, namely $e=(u_2-u_1)/(u_2+u_1)$. Based on the above analysis, the relations between the parameters are obtained as follows:
\begin{align}
\label{canshu}
u_2-u_1&=\frac{2e}{\lambda},
\\
\label{canshutwo}
u_3-u_1&=\frac{\left(1+g^2u_3^2\right)^{3/2}}{2M}+\frac{e-3}{\lambda}.
\end{align}
In these trajectories, the radial coordinate of the timelike test particle around the Bardeen black hole is bounded by $1/u_2\leq r \leq 1/u_1$. Therefore, 
\begin{align}
\label{drdubaone}
\phi(u_2)\simeq\frac{2}{\sqrt{\frac{2M}{(1+g^2u_2^2)^{3/2}}(u_3-u_1)}}K\left(\sqrt{\frac{u_2-u_1}{u_3-u_1}}\right),
\end{align}
where $K$ is the complete elliptic integral of the first kind. Using Eq.~(\ref{drdubaone}) and combining Eqs.~(\ref{rootone})--(\ref{rootthree}), the azimuthal angle can be further calculated as

\begin{align}
\label{drduba}
\begin{split}
\Delta\phi &= 2\phi(u_2)
\\
&\simeq \frac{4}{\sqrt{1+\frac{2M}{\lambda(1+g^2u_2^2)^{3/2}}(e-3)}}
\\
&\quad \times K\left(\sqrt{\frac{4Me}{\lambda(1+g^2u_2^2)^{3/2}+2M(e-3)}}\right).
\end{split}
\end{align}

\subsection{Periodicity Condition and $E$-$L$ Parametrization}
\label{sec:3.2}

Using the expression for $\Delta\phi$ obtained above, we define the parameter $q$ as
\begin{equation}
\label{drdujiu}
\begin{split}
q &= \frac{\Delta\phi}{2\pi}-1 
\\
&\simeq \frac{2}{\pi\sqrt{1+\frac{2M}{\lambda(1+g^2u_2^2)^{3/2}}(e-3)}} 
\\
&\quad \times K\left(\sqrt{\frac{4Me}{\lambda(1+g^2u_2^2)^{3/2}+2M(e-3)}}\right)-1.
\end{split}
\end{equation}
The parameter $q$ quantifies the deviation of the orbital period from a purely radial oscillation. For $q=0$, the orbit is a closed ellipse; for a rational number $q=w+v/z$, the orbit is periodic and characterized by the triplet $(z,w,v)$, where $z$, $w$ and $v$ denote the zoom number, whirl number, and vertex number, respectively~\cite{Levin2008PRD77.103005}. In practice, for a given $q$, the triplet $(z,w,v)$ can be obtained by truncating the continued fraction expansion of $q$ or by a direct numerical search over integer values of $z$, $w$ and $v$~\cite{Lim2024PRD109.024037,Chan2025GRG57.35,Huang2025PRD111.084038}.

We note that the dependence of $q$ on the energy $E$ and angular momentum $L$, as well as the locations of the MBO and the ISCO, has been analyzed in detail in~\cite{Gao2020AP418.168194}. Here we focus on the construction of periodic orbits from the orbital parameters $e$ and $\lambda$, and refer the reader to~\cite{Gao2020AP418.168194,Stuchlik2015IJMPD2450020} for the properties of the circular geodesic structure.

Setting $Q(u)=0$ in Eq.~(\ref{drdutt}) and solving the resulting cubic equation, the energy and angular momentum can be expressed in terms of the mass as
\begin{align}
\label{nlone}
\frac{1-E^2}{L^2}&=\frac{\left[(1+g^2u_2^2)^{3/2}\lambda-4M\right](1-e^2)}{\lambda^3(1+g^2u_2^2)^{3/2}},
\\
\label{jdlone}
\frac{M}{L^2}&=\frac{\lambda(1+g^2u_2^2)^{3/2}-3M-Me^2}{\lambda^2}.
\end{align}
The energy and angular momentum are then calculated as
\begin{align}
\label{nltwo}
E&=\sqrt{\frac{\lambda^2(1+g^2u_2^2)^3-4M\lambda(1+g^2u_2^2)^{3/2}+4M^2(1-e^2)}{\lambda(1+g^2u_2^2)^{3/2}\left[\lambda(1+g^2u_2^2)^{3/2}-3M-Me^2\right]}},
\nonumber  \\
\\
\label{jdltwo}
L&=\lambda\sqrt{\frac{M}{\lambda(1+g^2u_2^2)^{3/2}-3M-Me^2}}.
\end{align}
Equations~(\ref{nltwo}) and (\ref{jdltwo}) express $E$ and $L$ as functions of $\lambda$ and $e$. Here, the geometric factor is evaluated at the turning point $u_2$ of the radial motion; an equivalent expression can be obtained by using $u_1$. For the case $g=0$, Eqs~(\ref{nltwo}) and (\ref{jdltwo}) reduce to 
\begin{eqnarray}
\label{nlthree}
E&=&\sqrt{\frac{\lambda^2-4M\lambda+4M^2-4Me^2}{\lambda\left(\lambda-3M-Me^2\right)}},
\\
\label{jdlthree}
L&=&\lambda\sqrt{\frac{M}{\lambda-3M-Me^2}}.
\end{eqnarray}
These expressions are consistent with those in Ref.~\cite{Lim2024PRD109.024037}. We now consider the additional condition required for the orbit to be periodic. For a periodic orbit characterized by $(z,w,v)$ and $e$, the latus rectum $\lambda$ is not arbitrary, but must satisfy the rationality condition $q=w+v/z$. 
In the next section, we describe how this condition can be used to determine $\lambda$ and construct periodic orbits with prescribed $(z,w,v;e)$.

\subsection{Constructing Periodic Orbits with Prescribed $(z,w,v;e)$}
\label{sec:3.3}

It follows from Eq.~(\ref{drdujiu}) that a periodic orbit requires
\begin{align}
\label{rat}
K\left(k\right)\simeq \mathcal{S}\cdot \pi\sqrt{1+\frac{2M}{\lambda(1+g^2u_2^2)^{3/2}}(e-3)},
\end{align}
where $k=\sqrt{\frac{4Me}{\lambda(1+g^2u_2^2)^{3/2}+2M(e-3)}}$, and $\mathcal{S}=(q+1)/2$. The orbit is periodic when $q$ is a rational number of the form $q=w+v/z$. Equivalently, the rationality condition provides a relation between $\lambda$ and $e$ for a fixed $(z,w,v)$.

Imposing the periodicity condition $q=w+v/z$ in Eq.~(\ref{rat}), the corresponding value of $\lambda$ can be determined for a desired periodic orbit $(z,w,v)$ with orbital eccentricity $e$. Equations~(\ref{nltwo}) and (\ref{jdltwo}) then give the corresponding energy $E$ and angular momentum $L$. A key feature of this method is that  $e$, together with $(z,w,v)$, determines $\lambda$, and hence $E$ and $L$. The procedure for obtaining a periodic orbit around a Bardeen black hole is as follows:
\begin{enumerate}
\item Specify the periodic orbit by $(z,w,v)$ together with its orbital eccentricity $e$. The rational number $q$ is then calculated as $q=w+v/z$.
\item Use the obtained values of $q$ and $e$ to solve for $\lambda$ from Eq.~(\ref{rat}) [or equivalently from Eq.~(\ref{drdujiu})] using a numerical root-finding method.
\item Finally, the corresponding physical parameters, namely the energy $E$ and angular momentum $L$, are determined from Eqs.~(\ref{nltwo}) and (\ref{jdltwo}) after $\lambda$ has been obtained.
\end{enumerate}
This procedure allows us to construct periodic orbits with the desired $(z,w,v;e)$ classification in the Bardeen black hole spacetime.

\begin{figure*}
\includegraphics[width=0.95\textwidth]{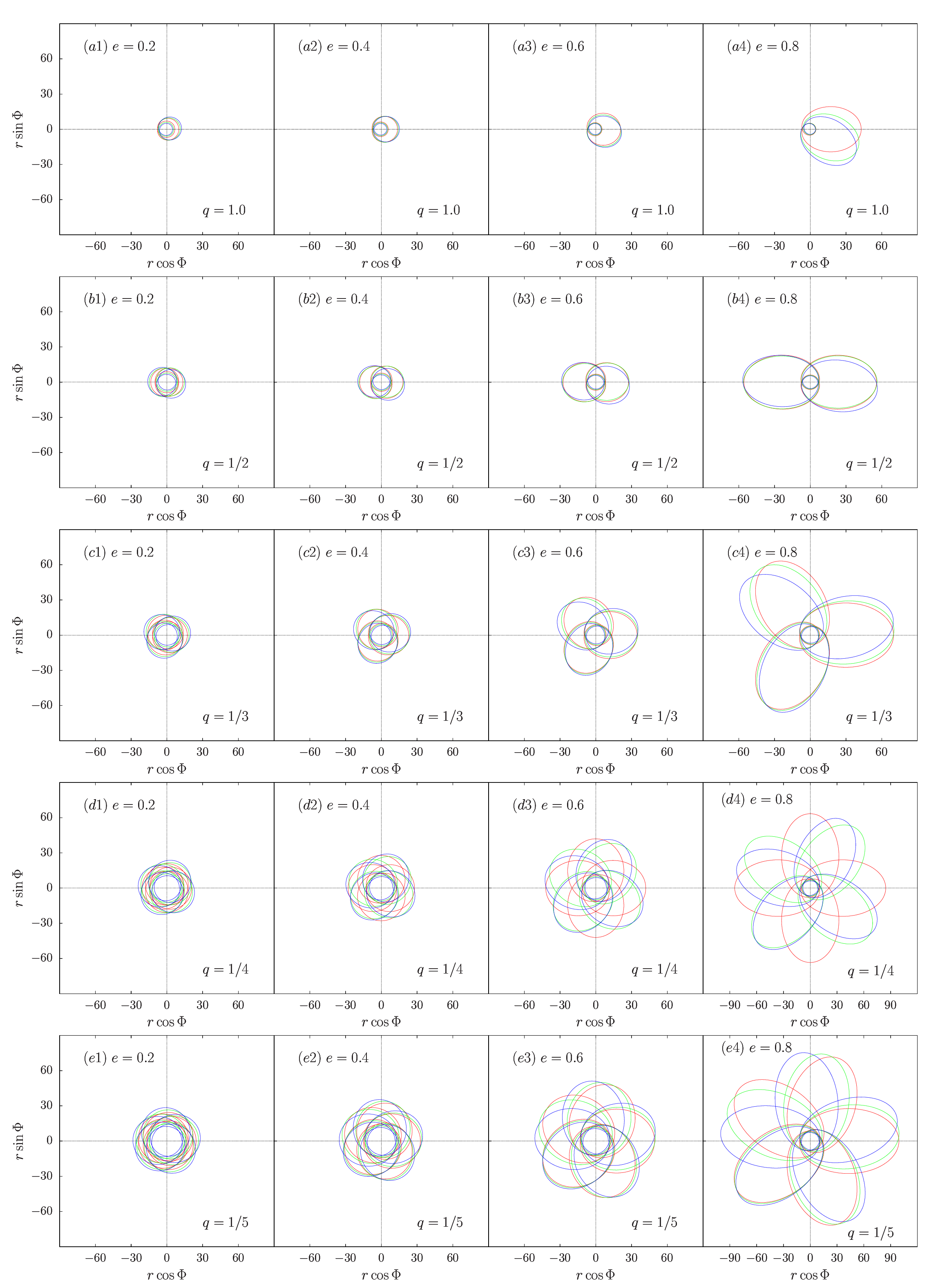}
\caption{Periodic orbits of the Bardeen black hole for different values of $q$ and $e$. The orbits are characterized by $(z,w,v)$ with $w=0$ and $v=1$; the corresponding values of $z$ are determined by $q=w+v/z$. The curves correspond to $g=0$ (red), $g=\sqrt{4/27}M$ (green), and $g=\sqrt{16/27}M$ (blue). 
\label{fig:2}}
\end{figure*}

\begin{table*}
\caption{Energy $E$, angular momentum $L$, and latus rectum $\lambda$ for the $q$ and $e$ values shown in Fig.~\ref{fig:2} with $g=0$. In this case, the Bardeen black hole reduces to the Schwarzschild black hole.}
\label{tab:1}
\begin{tabular*}{\textwidth}{@{\extracolsep{\fill}}lrrrrl@{}}
\hline
$q$& \multicolumn{1}{c}{$e=0.2$} & \multicolumn{1}{c}{$e=0.4$} & \multicolumn{1}{c}{$e=0.6$} & \multicolumn{1}{c}{$e=0.8$} \\
\hline
{       }& $\lambda=8.039631722M$ & $\lambda=8.154403633M$ & $\lambda=8.333799571M$&  $\lambda=8.564826709M$ \\
{$q=1.0$}& $L=3.595565028M$ & $L=3.648802743M$ & $L=3.736791885M$ & $L=3.859429425M$   \\
{       }& $E=0.9505365103$ & $E=0.9561974448$ & $E=0.9659637298$ & $E=0.9803265534$    \\
\hline
{       }& $\lambda=10.822436393M$ & $\lambda=10.888999038M$ & $\lambda=10.997566474M$&  $\lambda=11.14495580M$ \\
{$q=1/2$}& $L=3.879424035M$ & $L=3.916754335M$ & $L=3.979413731M$ & $L=4.068218600M$   \\
{       }& $E=0.9603319839$ & $E=0.9650087294$ & $E=0.9729757493$ & $E=0.9845038572$  \\
\hline
{       }& $\lambda=13.732035308M$ & $\lambda=13.784950102M$ & $\lambda=13.872058683M$&  $\lambda=13.991837594M$ \\
{$q=1/3$}& $L=4.199568981M$ & $L=4.229039570M$ & $L=4.278554112M$ & $L=4.348765910M$   \\
{       }& $E=0.9676607919$ & $E=0.9715356568$ & $E=0.9780966011$ & $E=0.9875046770$   \\
\hline
{       }& $\lambda=16.682274652M$ & $\lambda=16.728896191M$ & $\lambda=16.805936910M$&  $\lambda=16.912447250M$ \\
{$q=1/4$}& $L=4.516602522M$ & $L=4.541455115M$ & $L=4.583182957M$ & $L=4.642275934M$   \\
{       }& $E=0.9728840535$ & $E=0.9761638964$ & $E=0.9816979357$ & $E=0.9895914630$    \\
\hline
{       }& $\lambda=19.650751740M$ & $\lambda=19.693774243M$ & $\lambda=19.765012560M$&  $\lambda=19.863791435M$ \\
{$q=1/5$}& $L=4.821526136M$ & $L=4.843317652M$ & $L=4.879876280M$ & $L=4.931578713M$   \\
{       }& $E=0.9767140408$ & $E=0.9795478141$ & $E=0.9843184400$ & $E=0.9910997548$   \\
\hline
\end{tabular*}
\end{table*}

\begin{table*}
\caption{Energy $E$, angular momentum $L$, and latus rectum $\lambda$ for the $q$ and $e$ values shown in Fig.~\ref{fig:2}, with $g=\sqrt{4/27}M$.}
\label{tab:2}
\begin{tabular*}{\textwidth}{@{\extracolsep{\fill}}lrrrrl@{}}
\hline
$q$& \multicolumn{1}{c}{$e=0.2$} & \multicolumn{1}{c}{$e=0.4$} & \multicolumn{1}{c}{$e=0.6$} & \multicolumn{1}{c}{$e=0.8$} \\
\hline
{       }& $\lambda=7.958216282M$ & $\lambda=8.071825924M$ & $\lambda=8.249405163M$&  $\lambda=8.478092743M$ \\
{$q=1.0$}& $L=3.559153595M$ & $L=3.611852185M$ & $L=3.698950281M$ & $L=3.820345899M$   \\
{       }& $E=0.9505365103$ & $E=0.9561974448$ & $E=0.9659637298$ & $E=0.9803265534$    \\
\hline
{       }& $\lambda=10.712840151M$ & $\lambda=10.778728732M$ & $\lambda=10.886196731M$&  $\lambda=11.032093481M$ \\
{$q=1/2$}& $L=3.840138029M$ & $L=3.877090295M$ & $L=3.939115153M$ & $L=4.027020716M$   \\
{       }& $E=0.9603319839$ & $E=0.9650087294$ & $E=0.9729757493$ & $E=0.9845038572$  \\
\hline
{       }& $\lambda=13.592974249M$ & $\lambda=13.645353186M$ & $\lambda=13.731579640M$&  $\lambda=13.850145578M$ \\
{$q=1/3$}& $L=4.157040944M$ & $L=4.186213092M$ & $L=4.235226211M$ & $L=4.304726990M$   \\
{       }& $E=0.9676607919$ & $E=0.9715356568$ & $E=0.9780966011$ & $E=0.9875046766$   \\
\hline
{       }& $\lambda=16.513337219M$ & $\lambda=16.559486633$ & $\lambda=16.635747179M$&  $\lambda=16.741178914M$ \\
{$q=1/4$}& $L=4.470863961M$ & $L=4.495464878M$ & $L=4.536770152M$ & $L=4.595264708M$   \\
{       }& $E=0.9728840535$ & $E=0.9761638964$ & $E=0.9816979356$ & $E=0.9895914630$    \\
\hline
{       }& $\lambda=19.451753246M$ & $\lambda=19.494340069M$ & $\lambda=19.564857012M$&  $\lambda=19.662635537M$ \\
{$q=1/5$}& $L=4.772699686M$ & $L=4.794270525M$ & $L=4.830458931M$ & $L=4.881637787M$   \\
{       }& $E=0.9767140408$ & $E=0.9795478141$ & $E=0.9843184400$ & $E=0.9910997548$   \\
\hline
\end{tabular*}
\end{table*}

\begin{table*}
\caption{Energy $E$, angular momentum $L$, and latus rectum $\lambda$ for the $q$ and $e$ values shown in Fig.~\ref{fig:2}, with $g=\sqrt{16/27}M$.}
\label{tab:3}
\begin{tabular*}{\textwidth}{@{\extracolsep{\fill}}lrrrrl@{}}
\hline
$q$ & $e=0.2$ & $e=0.4$ & $e=0.6$ & $e=0.8$ \\
\hline
{       }& $\lambda=7.642318027M$ & $\lambda=7.751417979M$ & $\lambda=7.921948280M$&  $\lambda=8.141558197M$ \\
{$q=1.0$}& $L=3.417874398M$ & $L=3.468481138M$ & $L=3.552121910M$ & $L=3.668698777M$   \\
{       }& $E=0.9505365103$ & $E=0.9561974448$ & $E=0.9659637298$ & $E=0.9803265534$    \\
\hline
{       }& $\lambda=10.28759819M$ & $\lambda=10.350871349M$ & $\lambda=10.454073448M$&  $\lambda=10.594178885M$ \\
{$q=1/2$}& $L=3.687705265M$ & $L=3.723190726M$ & $L=3.782753532M$ & $L=3.867169718M$   \\
{       }& $E=0.9603319839$ & $E=0.9650087294$ & $E=0.9729757493$ & $E=0.9845038572$  \\
\hline
{       }& $\lambda=13.05340650M$ & $\lambda=13.103706277M$ & $\lambda=13.186510005M$&  $\lambda=13.300369515M$ \\
{$q=1/3$}& $L=3.992028844M$ & $L=4.0200430151M$ & $L=4.067110578M$ & $L=4.133852551M$   \\
{       }& $E=0.9676607919$ & $E=0.9715356568$ & $E=0.9780966011$ & $E=0.9875046766$   \\
\hline
{       }& $\lambda=15.857846816M$ & $\lambda=15.902164347M$ & $\lambda=15.975397761M$&  $\lambda=16.076644424M$ \\
{$q=1/4$}& $L=4.293394780M$ & $L=4.317019174M$ & $L=4.356684851M$ & $L=4.412857489M$   \\
{       }& $E=0.9728840535$ & $E=0.9761638964$ & $E=0.9816979357$ & $E=0.9895914630$    \\
\hline
{       }& $\lambda=18.67962358M$ & $\lambda=18.720519934M$ & $\lambda=18.788237734M$&  $\lambda=18.882134980M$ \\
{$q=1/5$}& $L=4.583249256M$ & $L=4.603963848M$ & $L=4.638715771M$ & $L=4.687863101M$   \\
{       }& $E=0.9767140408$ & $E=0.9795478141$ & $E=0.9843184400$ & $E=0.9910997548$   \\
\hline
\end{tabular*}
\end{table*}

In this approach, each pair $(e,\lambda)$ determines a periodic orbit. Figure~\ref{fig:2} compares the projected orbits for three values of the magnetic charge parameter,
$g=0$, $g=\sqrt{4/27}M$, and $g=\sqrt{16/27}M$, using five values of $q$ ($q=1$, $1/2$, $1/3$, $1/4$, $1/5$) and four values of the orbital eccentricity, $e=0.2$, $0.4$, $0.6$, and $e=0.8$. The figure is organized as a $5 \times 4$ grid: each row corresponds to a fixed $q$, and each column to a fixed $e$. The panels are labeled $(a1)-(a4)$, $(b1)-(b4)$, $(c1)-(c4)$, $(d1)-(d4)$, $(e1)-(e4)$, with rows from top to bottom corresponding to decreasing $q$. In each panel, the red, green, and blue curves correspond to $g=0$, $g=\sqrt{4/27}M$, and $g=\sqrt{16/27}M$, respectively. The horizontal and vertical axes are $r\cos{\phi}$ and $r\sin{\phi}$, representing the projection of each trajectory onto the equatorial plane. Table~\ref{tab:1} lists the corresponding energy $E$, angular momentum $L$, and latus rectum $\lambda$ for $g=0$. Tables~\ref{tab:2} and~\ref{tab:3} list the same quantities for $g=\sqrt{4/27}M$ and $g=\sqrt{16/27}M$, respectively.

For the orbits shown in Fig.~\ref{fig:2}, the rational number $q$ takes the form $q=1/z$ with $z=1$, $2$, $3$, $4$, and $5$, corresponding to $(z,w,v)=(z,0,1)$. At fixed $q$, the values of $\lambda$, $E$, and $L$ increase with orbital eccentricity in the parameter ranges considered here. 
The red curves ($g=0$) reproduce the Schwarzschild black hole results. The figure also shows that, for fixed $q$ and $e$, increasing $g$ modifies the orbital shape, typically making the trajectory more distorted and increasing the number of self-intersections, especially at larger $e$. Decreasing $q$ from $1$ to $1/5$ increases the number of orbital zooms from one to five, consistent with the frequency ratio $\omega_{\phi}/\omega_{r}=1+q$ for a rational number $q=w+v/z$~\cite{Grossman2012PRD85.023012}.

The physical significance of these morphological changes is closely tied to the detectability of GWs from periodic orbits around Bardeen black holes. The periodicity of the orbital motion translates directly into the periodicity of GW emission, so that a periodic orbit generates a GW signal with discrete frequencies determined by the orbital harmonics. When $q$ is rational, the orbit closes after a finite number of cycles, producing a discrete spectrum of GW frequencies. A more compact or more distorted orbit with higher self-intersections, such as those found at larger $g$, radiates GWs with higher-frequency harmonics and may enhance the power in specific frequency bands. Figure~\ref{fig:2} therefore serves as a visual catalog from which GW templates for Bardeen black holes can be constructed, providing a basis for comparison with signals from the LVK network~\cite{Abbott2023PRX13.041039,2021arXiv211206861T} or future space-based detectors such as LISA~\cite{Robson2019CQG36.105011}. In this sense, the figure bridges the theoretical framework of regular black holes with the observational frontier of GW astronomy, establishing the foundation for computing and potentially detecting GW signatures from periodic orbits around Bardeen black holes.

We emphasize that the orbital shapes shown in Fig.~\ref{fig:2} are obtained by numerically integrating the exact geodesic equation with the parameter sets determined from the periodicity condition. The analytical elliptic-integral approximation is used only to solve for $\lambda$; it does not affect the accuracy of the orbital shapes.

\section{Gravitational Waves Corresponding to Periodic Orbits}
\label{sec:4}

The previous section described how the rational $q$, orbital eccentricity $e$, and magnetic charge parameter $g$ determine the energy and angular momentum of periodic orbits classified by $(z,w,v)$. Such periodic orbit classifications provide a useful framework for characterizing bound timelike geodesics around black holes~\cite{Levin2008PRD77.103005}. In this section, we investigate the GW signatures associated with these periodic orbits around a Bardeen black hole. This analysis characterizes the dynamics of timelike test particles on periodic orbits and the gravitational radiation they generate~\cite{Glampedakis2002PRD66.044002}.

\subsection{Time domain approach}
\label{sec:4.1}

In a strong gravitational field, bound timelike test particle orbits around a compact object may exhibit periodic or quasiperiodic motion~\cite{Deng2021PDU31.100745,Deng2023EPJC83.311,Huang2024PRD109.124005,Huang2025PRD111.084038,Huang2025JCAP10.067}. For periodic orbits, the classification proposed by Levin and Perez-Giz~\cite{Levin2008PRD77.103005} labels each orbit by three integers $(z,w,v)$, which determine the rational number $q=w+v/z$. Quasiperiodic orbits, in contrast, correspond to irrational values of $q$. Once the orbital motion is specified, the associated GW signal can be constructed using the numerical kludge waveform model, which has been shown to be effective for modeling waveforms from EMRIs~\cite{Babak2007PRD75.024005}. In the transverse traceless gauge, the two transverse traceless tensor polarizations, $h_{+}$ and $h_{\times}$, are given by~\cite{Poisson2014Book}
\begin{align}
\label{hplus}
h_{+}&=-\frac{2\eta}{D_{L}}\frac{G^2M^2}{c^4r}(1+{\cos}^2{\iota})\cos{(2\phi+2\zeta)},\\
\label{hcross}
h_{\times}&=-\frac{4\eta}{D_{L}}\frac{G^2M^2}{c^4r}{\cos}{\iota}\sin{(2\phi+2\zeta)}.
\end{align}
Here, $m$ represents the mass of the small object (e.g., a timelike test particle), $M$ denotes the mass of the supermassive object (such as a Bardeen black hole), $D_{L}$ is the luminosity distance, $\iota$ is the inclination angle, and $\zeta$ is the longitude of periastron. Let $\phi$ denote the phase angle, and let $\eta=mM/(m+M)^2$ denote the symmetric mass ratio. In Eqs.~(\ref{hplus}) and (\ref{hcross}), $r$ is the orbital radius of the timelike test particle.

\begin{figure*}
\includegraphics[width=1.0\textwidth]{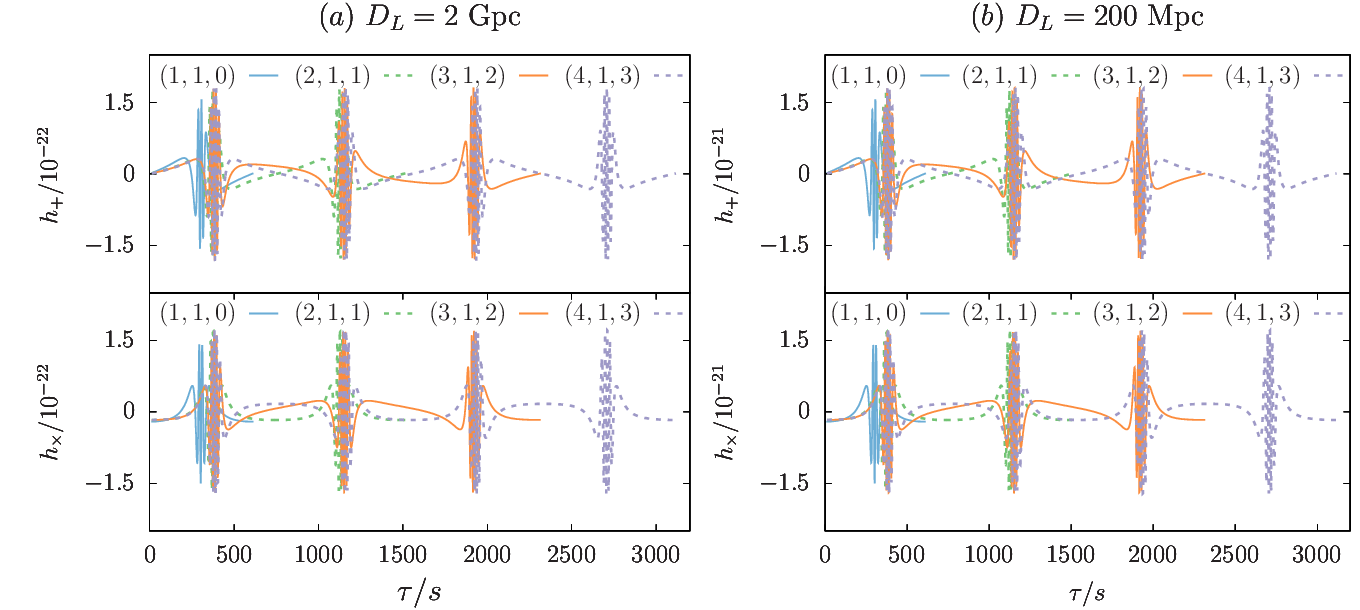}
\caption{Gravitational wave waveforms generated by the periodic orbits $(1,1,0)$, $(2,1,1)$, $(3,1,2)$, and $(4,1,3)$ are shown for fixed parameters $m=10M_{\odot}$, $M=10^6M_{\odot}$, and $\iota=\zeta=\pi/4$, and varying luminosity distance $D_L$. Panel $(a)$ corresponds to $D_L=2\,\mathrm{Gpc}$, and panel $(b)$ to $D_L=200\,\mathrm{Mpc}$. For the angular momentum $L=3.720353$, the energies of these four orbits are $E=0.975374$, $0.978592$, $0.978838$, and $0.978912$, respectively.
\label{fig:3}}
\end{figure*}

Since multiple parameter combinations can arise in practical applications, we consider several representative parameter sets. Our comparison focuses primarily on the effects of the supermassive black hole mass in EMRIs, as well as on those of the luminosity distance, inclination angle, and longitude of periastron. We calculate the two tensor polarizations, given in Eqs.~(\ref{hplus}) and (\ref{hcross}), for GWs generated by timelike test particles with mass $m=10M_{\odot}$ following different periodic orbits $(z,1,v)$ around Bardeen black holes. The Bardeen black hole parameter is fixed at $g=\sqrt{8/27}M$. 

Figure~\ref{fig:3} shows the GW waveforms associated with the periodic orbits $(z,1,v)$ for fixed parameters $m=10M_{\odot}$, $M=10^6M_{\odot}$, and $\iota=\zeta=\pi/4$, while varying the luminosity distance $D_L$. For the angular momentum $L=3.720353$, the energies of the periodic orbits $(1,1,0)$, $(2,1,1)$, $(3,1,2)$, and $(4,1,3)$ are $E=0.975374$, $0.978592$, $0.978838$, and $0.978912$, respectively. Since the orbital equations are written in geometric units and do not involve $D_L$, the time evolution of the waveform is identical in the two panels. As shown in Fig.~\ref{fig:3}, the luminosity distance has a significant effect on the GW strain amplitudes. In panel~$(a)$, where $D_L=2\,\mathrm{Gpc}$, the strain amplitudes are of order $10^{-22}$. In panel~$(b)$, where $D_L=200\,\mathrm{Mpc}$, the strain amplitudes are of order $10^{-21}$. Thus, decreasing $D_L$ by one order of magnitude increases the strain amplitude by about one order of magnitude, consistent with $h \propto 1/D_L$.

\begin{figure*}
\includegraphics[width=1.0\textwidth]{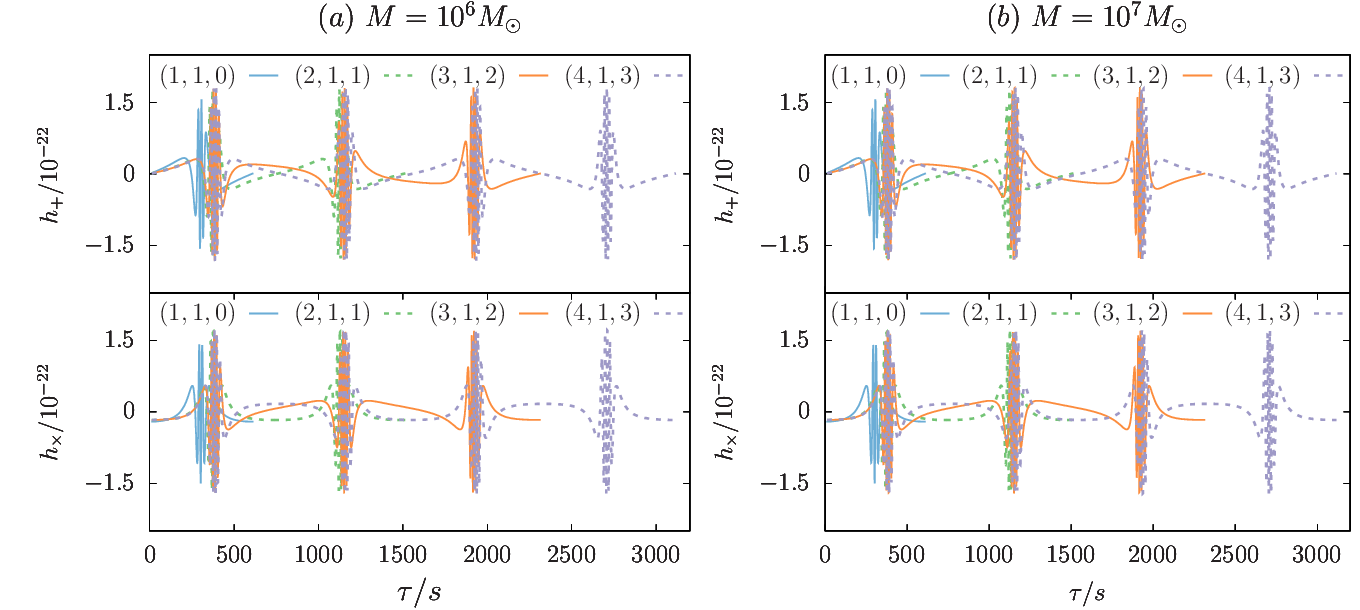}
\caption{Gravitational wave waveforms generated by the periodic orbits $(1,1,0)$,
$(2,1,1)$, $(3,1,2)$, and $(4,1,3)$ for fixed parameters
$m=10M_{\odot}$, $D_L=2\,\mathrm{Gpc}$, and
$\iota=\zeta=\pi/4$, with varying central black hole mass $M$. Panel $(a)$ corresponds to $M=10^6M_{\odot}$, and panel $(b)$ to $M=10^7M_{\odot}$. The energies and the angular momentum are the same as those in Fig.~\ref{fig:3}.
\label{fig:4}}
\end{figure*}

Figure~\ref{fig:4} shows the GW waveforms associated with the periodic orbits $(z,1,v)$ for fixed parameters $m=10M_{\odot}$, $D_L=2\,\mathrm{Gpc}$, and $\iota=\zeta=\pi/4$, while varying the mass of the central black hole $M$. The black hole masses are $M=10^6M_{\odot}$ in panel~$(a)$ and $M=10^7M_{\odot}$ in panel~$(b)$. 
Because the orbital dynamics are computed in geometric units with a fixed dimensionless angular momentum $L$ and energy $E$, the functions $u(\phi)$ and $\tau(\phi)$ are independent of $M$. Consequently, the two panels display exactly the same waveform shape and time structure.
A comparison of the two panels shows that the strain amplitudes and the visible waveform time structure remain almost unchanged. This is consistent with the amplitude scaling for $m\ll M$, $\eta \approx m/M$, and with a geometrically scaled orbital radius $r\propto GM/c^2$, the strain amplitude approximately satisfies $h\propto m$, independent of $M$. The energies and the angular momentum used in Fig.~\ref{fig:4} are the same as those in Fig.~\ref{fig:3}.

\begin{figure*}
\includegraphics[width=1.0\textwidth]{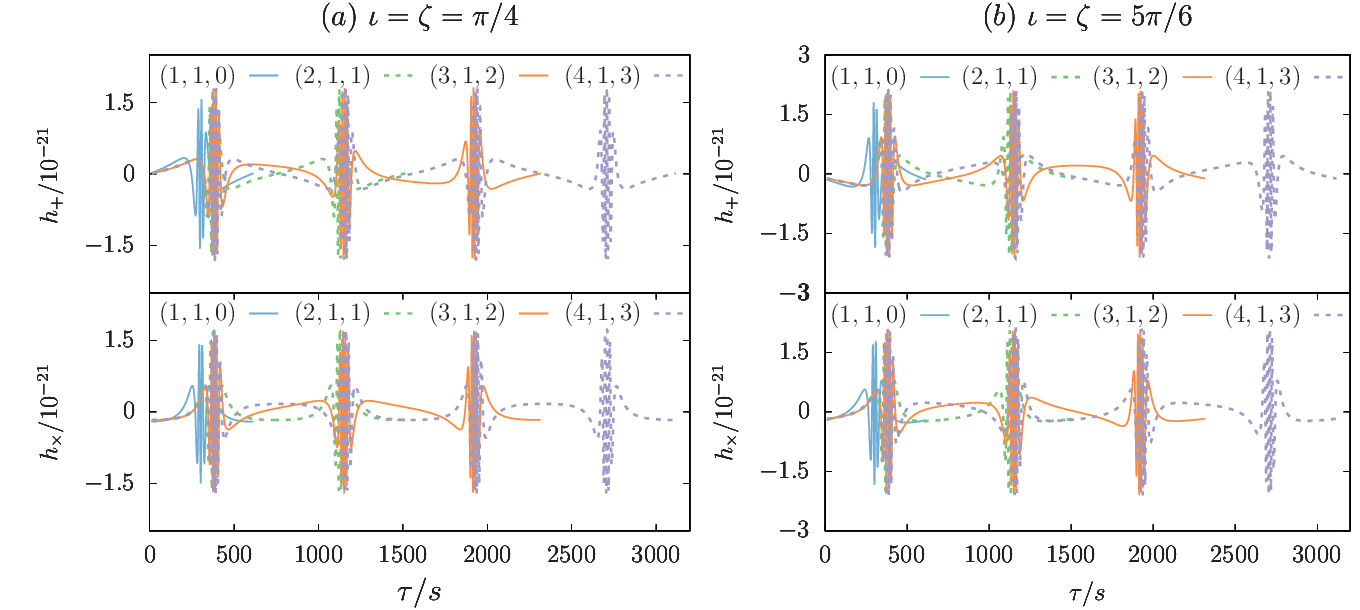}
\caption{Gravitational wave waveforms generated by the periodic orbits $(1,1,0)$, $(2,1,1)$, $(3,1,2)$, and $(4,1,3)$ for fixed parameters $m=10M_{\odot}$, $M=10^6M_{\odot}$, and
$D_L=200\,\mathrm{Mpc}$, with varying inclination angle $\iota$ and longitude of periastron $\zeta$. Panel $(a)$ corresponds to $\iota=\zeta=\pi/4$, and panel $(b)$ to $\iota=\zeta=5\pi/6$. The energies and the angular momentum are the same as those in Fig.~\ref{fig:3}.
\label{fig:5}}
\end{figure*}

Figure~\ref{fig:5} presents the GW waveforms for fixed parameters $m=10M_{\odot}$, $M=10^6M_{\odot}$, and $D_L=200\,\mathrm{Mpc}$, while varying the inclination angle $\iota$ and the longitude of periastron $\zeta$. In panel~$(a)$, $\iota=\zeta=\pi/4$, whereas in panel~$(b)$, $\iota=\zeta=5\pi/6$. The strain amplitudes remain of the same order in both panels, while the waveform envelope and phase structure change with $\iota$ and $\zeta$. This is consistent with Eqs.~(\ref{hplus}) and (\ref{hcross}), $\iota$ enters only through the amplitude coefficients $(1+\cos^2\iota)$ and $\cos\iota$, while $\zeta$ appears only as a phase shift $2\zeta$ in the arguments of the trigonometric functions. Thus, variations in $\iota$ and $\zeta$ mainly modify the waveform details and phase structure, leaving the overall strain amplitude at the same order of magnitude. The energies and the angular momentum used in Fig.~\ref{fig:5} are the same as those in Fig.~\ref{fig:3}.

Figures~\ref{fig:3}--\ref{fig:5} present the numerically computed GW waveforms in the time domain. The transverse traceless tensor polarizations, $h_{+}$ and $h_{\times}$, are plotted as functions of the proper time $\tau$ measured in seconds. The blue, green, orange, and light-purple curves correspond to the periodic orbits $(1,1,0)$, $(2,1,1)$, $(3,1,2)$, and $(4,1,3)$, respectively. The results show that the luminosity distance $D_L$ has a significant effect on the waveform amplitude, whereas the central black hole mass $M$, inclination angle $\iota$, and longitude of
periastron $\zeta$ have comparatively minor effects on the amplitude. The two tensor polarizations also exhibit distinct phase evolution and clearly reveal the zoom-whirl structure of the GWs emitted by a Bardeen black hole. In particular, both polarizations display the zoom-whirl behavior associated with the corresponding periodic orbits: the number of zooms in an orbit is reflected in the number of distinct phase features in the GW signal, whereas the whirls are associated with the higher-amplitude peaks.

\subsection{Frequency domain approach}
\label{sec:4.2}

In astronomical research, time-varying signals emitted by celestial bodies, such as luminosity and electromagnetic radiation intensity, can generally be analyzed by decomposing them into different frequency components. This decomposition facilitates the investigation and interpretation of the physical properties and dynamical laws of celestial objects. Similarly, the two tensor polarizations of GWs can be transformed from the time domain to the frequency domain using the Fourier transform. Although Fourier analysis involves a variety of sophisticated methods, the discrete Fourier transform (DFT) is one of the simplest. As a preliminary step before introducing the frequency domain analysis~\cite{Speri2023arXiv2307.12585}, this section assumes that the GW signals from EMRIs around a Bardeen black hole are discretely sampled and have finite duration. Under these assumptions, the DFT can be used to investigate the frequency domain properties of the two GW polarizations.

\begin{figure*}
\includegraphics[width=1.05\textwidth]{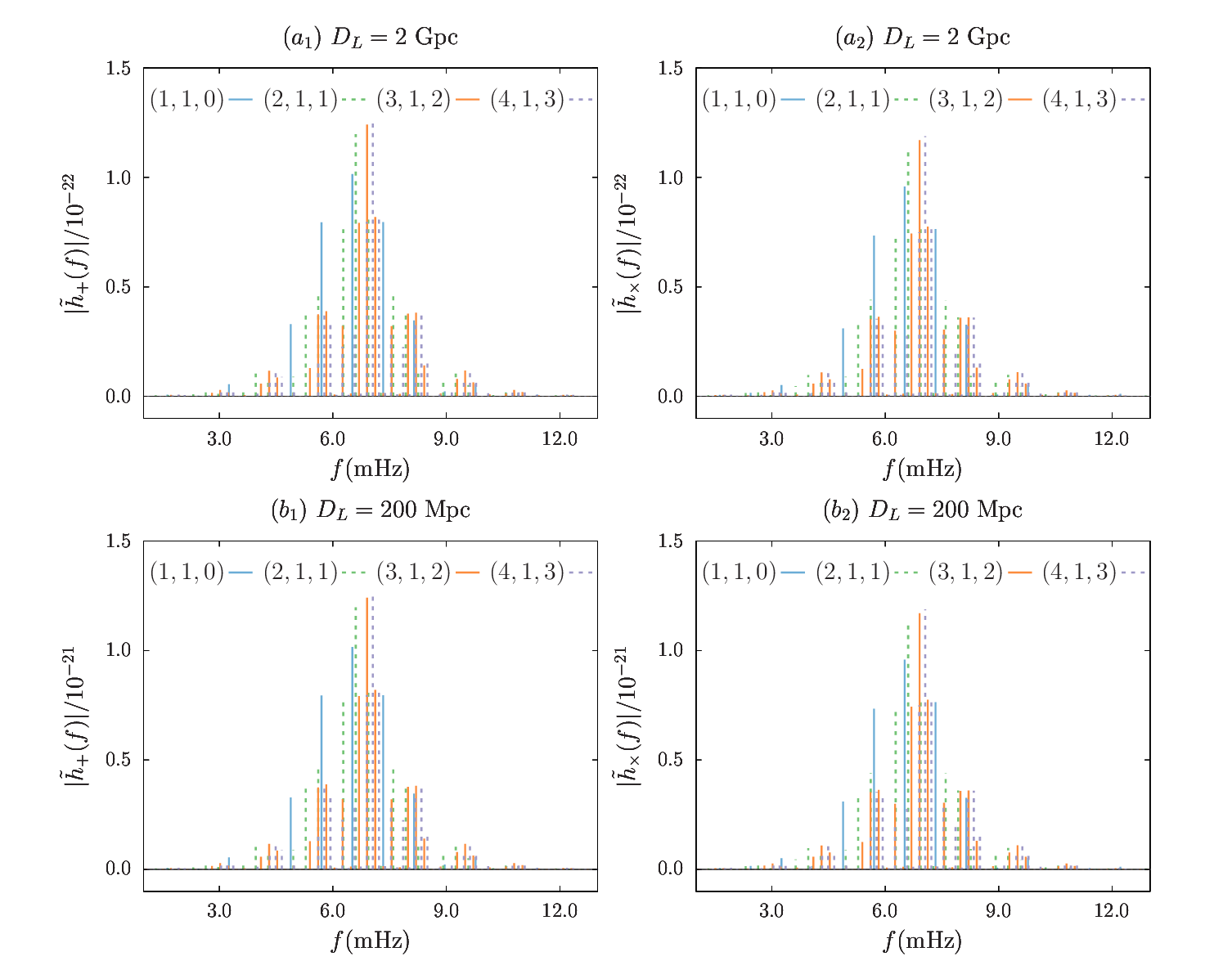}
\caption{Absolute values of the frequency domain gravitational wave waveforms, $|\tilde{h}_{+}(f)|$ and $|\tilde{h}_{\times}(f)|$, corresponding to the periodic orbits $(z,1,v)$, calculated with fixed $m=10M_{\odot}$, $M=10^6M_{\odot}$, and $\iota=\zeta=\pi/4$, while varying the luminosity distance $D_L$. Panels $(a1)$ and $(a2)$ correspond to $D_L=2\,\mathrm{Gpc}$, and panels $(b1)$ and $(b2)$ to $D_L=200\,\mathrm{Mpc}$. The energy and angular momentum values corresponding to the periodic orbits $(1,1,0)$, $(2,1,1)$, $(3,1,2)$, and $(4,1,3)$ are the same as those used in Fig.~\ref{fig:3}.
\label{fig:6}}
\end{figure*}

\begin{figure*}
\includegraphics[width=1.05\textwidth]{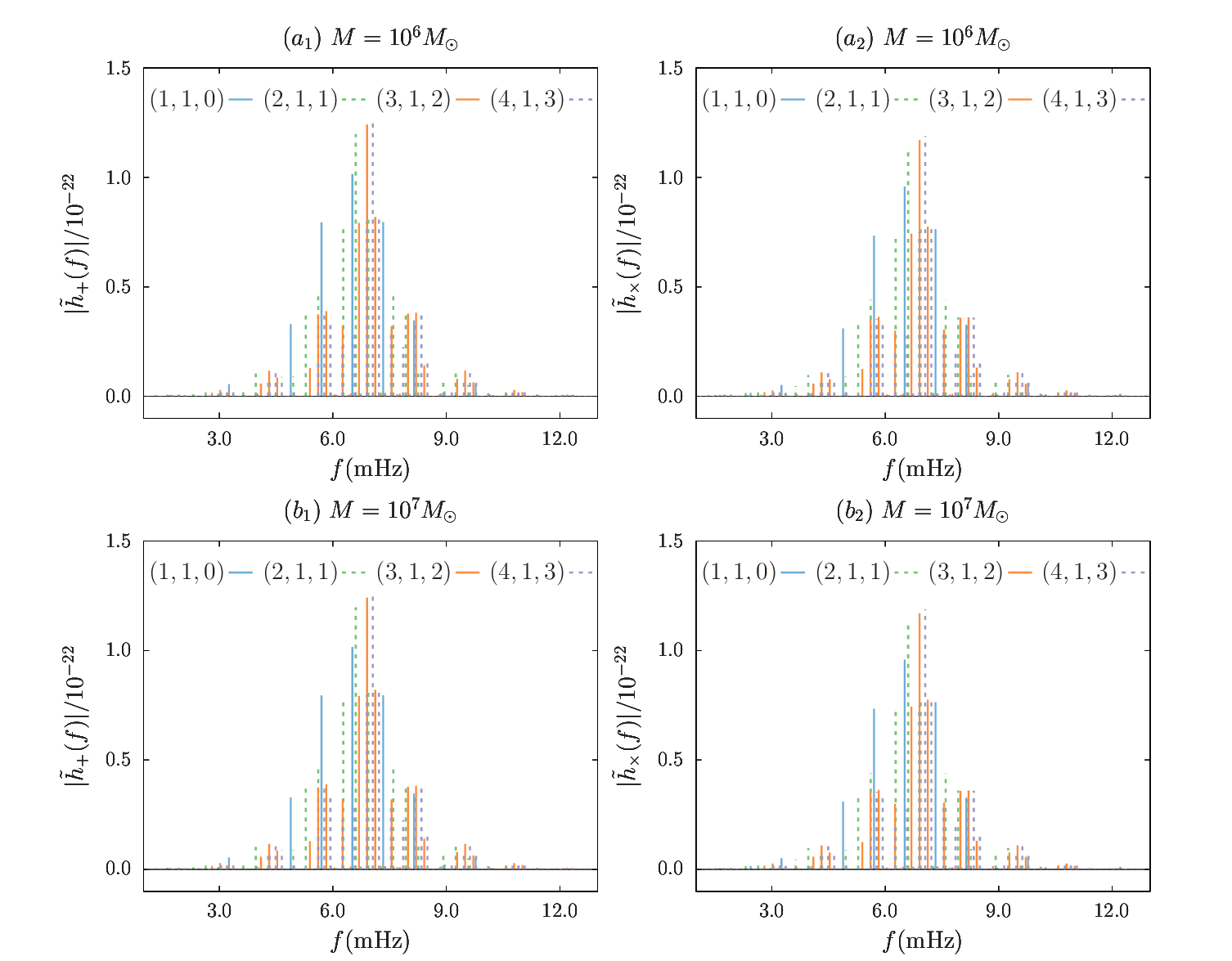}
\caption{Absolute values of the frequency domain gravitational wave waveforms, $|\tilde{h}_{+}(f)|$ and $|\tilde{h}_{\times}(f)|$, corresponding to the periodic orbits $(z,1,v)$, calculated with fixed $m=10M_{\odot}$, $D_L=2\,\mathrm{Gpc}$, and $\iota=\zeta=\pi/4$, while varying the central black hole mass $M$. Panels $(a1)$ and $(a2)$ correspond to $M=10^6M_{\odot}$, and panels $(b1)$ and $(b2)$ to $M=10^7M_{\odot}$. The energy and angular momentum values corresponding to the periodic orbits $(1,1,0)$, $(2,1,1)$, $(3,1,2)$, and $(4,1,3)$ are the same as those in Fig.~\ref{fig:4}.
\label{fig:7}}
\end{figure*}

\begin{figure*}
\includegraphics[width=1.05\textwidth]{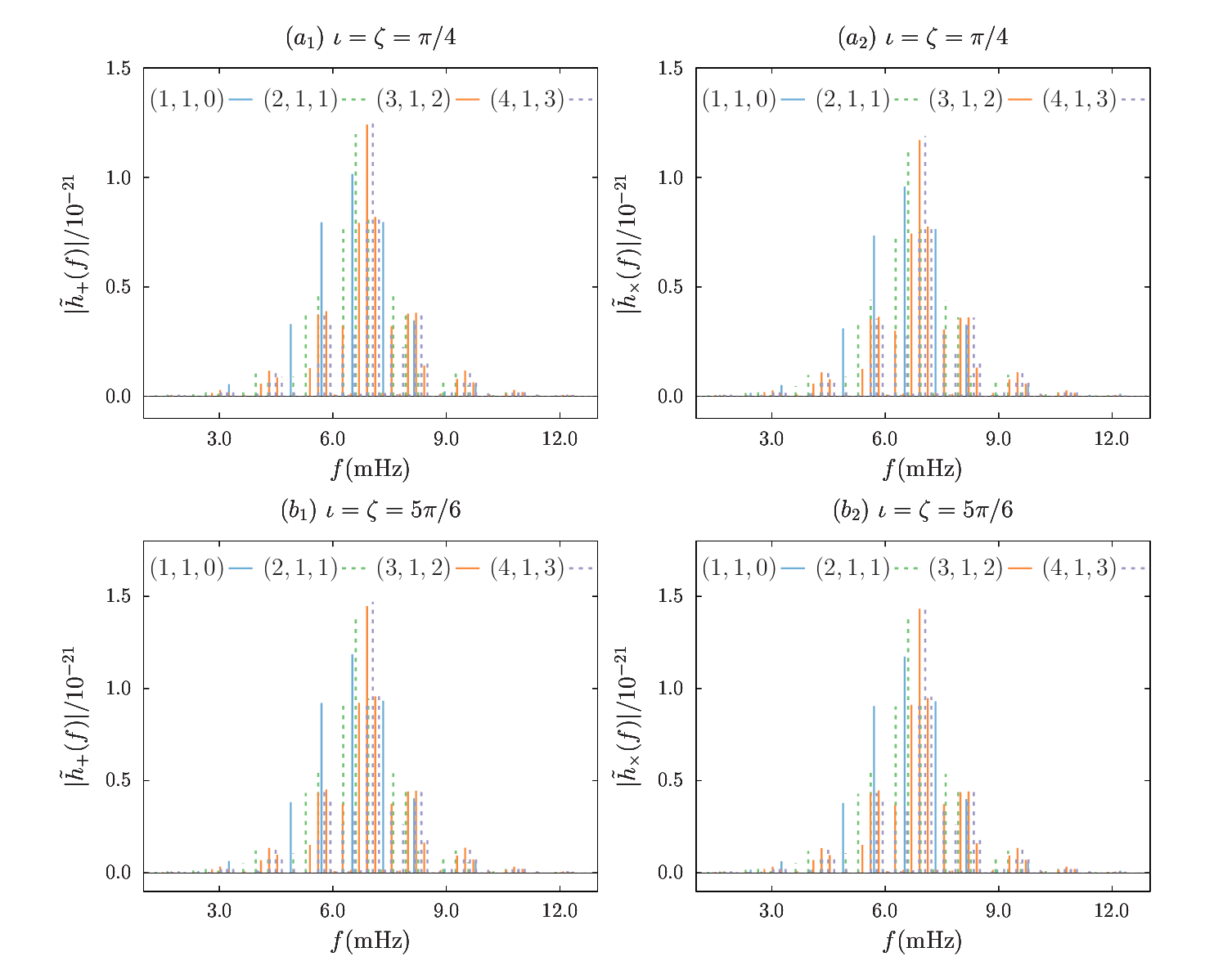}
\caption{Absolute values of the frequency domain gravitational wave waveforms, $|\tilde{h}_{+}(f)|$ and $|\tilde{h}_{\times}(f)|$, corresponding to the periodic orbits $(z,1,v)$, calculated with fixed $m=10M_{\odot}$, $M=10^6M_{\odot}$, and $D_L=200\,\mathrm{Mpc}$, while varying $\iota$ and $\zeta$. Panels $(a1)$ and $(a2)$ correspond to $\iota=\zeta=\pi/4$, and panels $(b1)$ and $(b2)$ to $\iota=\zeta=5\pi/6$. The energy and angular momentum values corresponding to the periodic orbits $(1,1,0)$, $(2,1,1)$, $(3,1,2)$, and $(4,1,3)$ are the same as those in Fig.~\ref{fig:5}.
\label{fig:8}}
\end{figure*}

By applying the DFT to the time domain GW waveforms shown in
Figs.~\ref{fig:3}, \ref{fig:4}, and \ref{fig:5}, we obtain the corresponding frequency domain waveforms for EMRIs on different periodic orbits around Bardeen black holes. Figures~\ref{fig:6}, \ref{fig:7}, and \ref{fig:8} show the absolute values of the two frequency domain GW polarizations, $|\tilde{h}_{+}(f)|$ and $|\tilde{h}_{\times}(f)|$. The results are presented for four different periodic orbits: $(1,1,0)$, $(2,1,1)$, $(3,1,2)$, and $(4,1,3)$. For clarity, the waveform amplitudes are scaled by factors of $10^{-22}$ and $10^{-21}$ for different parameter sets, as indicated by the vertical-axis labels, and the frequency is expressed in millihertz. Overall, the characteristic frequencies of the spectra fall within the sensitivity band of space-based laser interferometers, motivating a more detailed examination of the dependence of the frequency domain waveforms on the physical parameters.

Figure~\ref{fig:6} shows the frequency domain GW waveforms associated with the periodic orbits $(z,1,v)$ for fixed $m=10M_{\odot}$, $M=10^6M_{\odot}$, and $\iota=\zeta=\pi/4$, while varying the luminosity distance $D_L$. In panels $(a1)$ and $(a2)$, $D_L=2\,\mathrm{Gpc}$, whereas in panels $(b1)$ and $(b2)$, $D_L=200\,\mathrm{Mpc}$. Comparing the two rows, the peak frequencies and spectral shapes remain essentially unchanged, while the spectral amplitudes decrease by about one order of magnitude as $D_L$ increases by one order of magnitude, consistent with $h\propto 1/D_L$. Thus, $D_L$ mainly rescales the overall amplitude of the frequency domain waveform without shifting its characteristic frequencies or changing its spectral structure.

Figure~\ref{fig:7} shows the frequency domain GW waveforms associated with the periodic orbits $(z,1,v)$ for fixed parameters $m=10M_{\odot}$, $D_L=2\,\mathrm{Gpc}$, and $\iota=\zeta=\pi/4$, while varying the central black hole mass $M$. The black hole masses are $M=10^6M_{\odot}$ in panels $(a1)$ and $(a2)$ and $M=10^7M_{\odot}$ in panels $(b1)$ and $(b2)$. The spectra in the two rows are almost identical: the peak frequencies remain around $6$--$9\,\mathrm{mHz}$, and the amplitudes are of the same order. This is consistent with the time domain results in Fig.~\ref{fig:4}. For $m\ll M$, $\eta \approx m/M$, and with a geometrically scaled orbital radius $r\propto GM/c^2$, the strain amplitude approximately satisfies $h\propto m$, independent of $M$. The energy and angular momentum values used in Fig.~\ref{fig:7} are the same as those in Fig.~\ref{fig:4}.

Figure~\ref{fig:8} displays the frequency domain GW waveforms associated with the periodic orbits $(z,1,v)$ for fixed $m=10M_{\odot}$, $M=10^6M_{\odot}$, and
$D_L=200\,\mathrm{Mpc}$, while varying the inclination angle
$\iota$ and the longitude of periastron $\zeta$. In panels $(a1)$ and $(a2)$, $\iota=\zeta=\pi/4$, whereas in panels $(b1)$ and $(b2)$, $\iota=\zeta=5\pi/6$. The peak frequencies and overall amplitudes remain of the same order of magnitude in both cases, while the relative heights of the harmonics and the fine structure of the spectrum change with $\iota$ and $\zeta$. This is consistent with Eqs.~(\ref{hplus}) and (\ref{hcross}), $\iota$ enters only through the amplitude coefficients $(1+\cos^2\iota)$ and $\cos\iota$, while $\zeta$ appears only as a phase shift $2\zeta$ in the arguments of the trigonometric functions. Thus, variations in $\iota$ and $\zeta$ mainly modify the relative harmonic content and the spectral details, while leaving the overall amplitude and characteristic frequencies nearly unchanged.

In summary, the frequency domain results in Figs.~\ref{fig:6}--\ref{fig:8} mirror the time domain behavior in Figs.~\ref{fig:3}--\ref{fig:5}: the luminosity distance primarily rescales the overall amplitude of the spectrum, while the central black hole mass has only a weak effect on the spectrum for $m \ll M$, the inclination angle and longitude of periastron mainly affect the relative harmonic content and spectral details. These frequency domain results provide the characteristic strains used in the detector sensitivity analysis of the next subsection.

\subsection{Gravitational Wave Detection Prospects for EMRIs around Bardeen Black Holes}
\label{sec:4.3}

In GW astronomy, the sensitivity of a GW detector is commonly characterized by its sensitivity curve, which represents the detector's equivalent noise spectral density over a given frequency range. Depending on the convention, the vertical axis may represent the noise power spectral density $S_{\mathrm{n}}(f)$, the amplitude spectral density $\sqrt{S_{\mathrm{n}}(f)}$, or the characteristic strain $h_{\mathrm{c}}(f)$~\cite{Moore2015CQG32.015014,Robson2019CQG36.105011}.

When assessing the detectability of a potential GW source, the source's GW strain spectrum, usually expressed as its characteristic strain $h_{\mathrm{c},\mathrm{source}}(f)$,  should be compared with the detector's noise curve in the frequency domain~\cite{Moore2015CQG32.015014}. If the source's characteristic strain lies above the detector's characteristic noise strain over a sufficiently broad frequency range, the signal may be detectable using signal-processing techniques such as matched filtering~\cite{Allen2012PRD85.122006}. The actual detectability is determined by the integrated signal-to-noise ratio (SNR), which must exceed a predefined detection threshold~\cite{Berry2019BAAS51.42}. Conversely, if the source strain spectrum remains well below the detector's noise curve over the relevant frequency range, the signal may be difficult to distinguish from instrumental noise and may fail to satisfy the statistical detection criterion. Frequency domain comparisons therefore provide a useful preliminary method for evaluating the detectability of GW sources.

To assess the detectability of GW waveforms emitted by EMRIs on periodic orbits around Bardeen black holes, the characteristic strain can be computed as follows~\cite{Finn2000PRD62.l24021}:
\begin{eqnarray}
\label{tzyb}
h_{\mathrm{c}}(f)&=&2f\sqrt{\big|\tilde{h}_{+}(f)\big|^2+\big|\tilde{h}_{\times}(f)\big|^2}.
\end{eqnarray}
When assessing the detectability of these waveforms with current and future detectors, it is common to plot the characteristic strain curves $h_{\mathrm{c}}(f)$ of the expected signals and compare them with the sensitivity curves of various detectors. For EMRIs, we consider LISA~\cite{Robson2019CQG36.105011}, the evolved Laser Interferometer Space Antenna (eLISA)~\cite{eLISA2013arXiv1305.5720}, TianQin~\cite{Luo2016CQG33.035010}, BBO~\cite{Cutler2006PRD73.042001}, and DECIGO~\cite{DECIGO2021PTEP2021.05A105} as potential detectors. The characteristic strain is plotted as a function of frequency for various EMRI sources, together with the sensitivity curves of these detectors.

Building on the preceding analysis, we present the characteristic strain $h_{\mathrm{c}}(f)$ of GWs from different periodic orbits of EMRIs around Bardeen black holes as a function of frequency. To facilitate comparison, the sensitivity curves of the relevant space-based GW detectors are superimposed on the same axes.

\begin{figure*}
\includegraphics[width=1.025\textwidth]{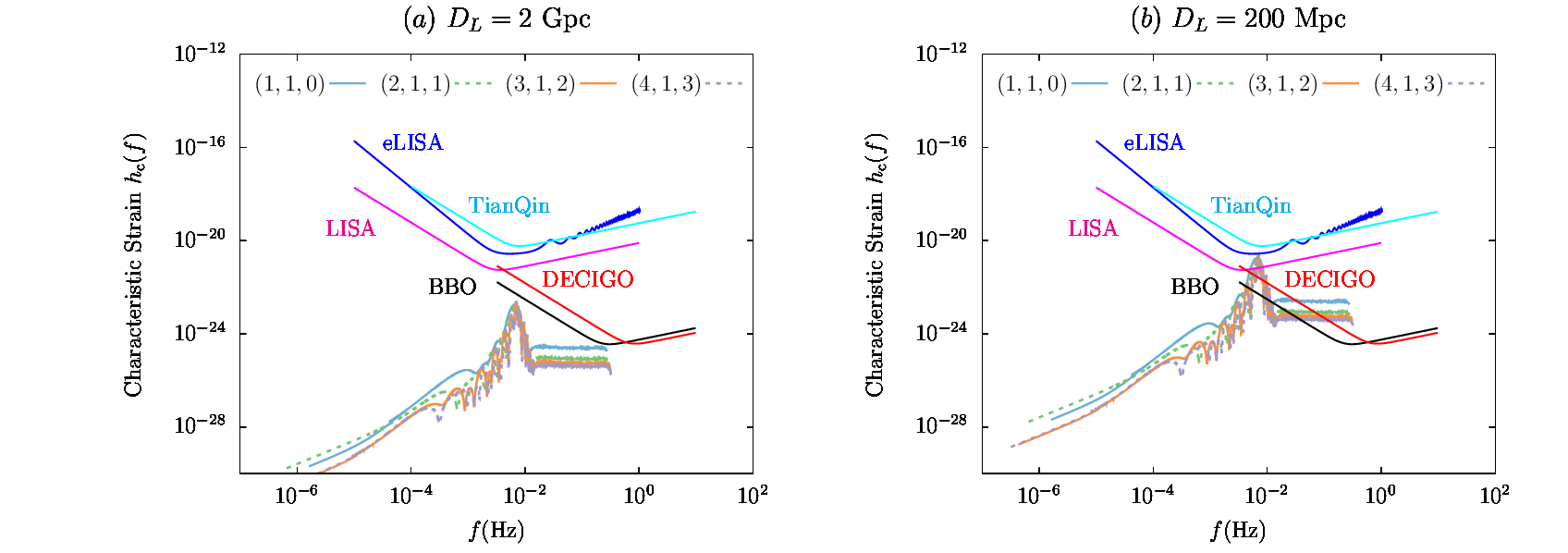}
\caption{Characteristic strain of gravitational waves from EMRIs on periodic orbits around Bardeen black holes, compared with the sensitivity curves of LISA (magenta), eLISA (blue), TianQin (cyan), BBO (black), and DECIGO (red). The left and right panels correspond to $D_L=2\,\mathrm{Gpc}$ and $D_L=200\,\mathrm{Mpc}$, respectively. The remaining parameters $m$, $M$, $\iota$, and $\zeta$ are fixed as in Fig.~\ref{fig:3}.
\label{fig:9}}
\end{figure*}

Figure~\ref{fig:9} shows the characteristic strain for fixed $m=10M_{\odot}$, $M=10^6M_{\odot}$, and $\iota=\zeta=\pi/4$, while varying the luminosity distance $D_L$. The left panel corresponds to $D_L=2\,\mathrm{Gpc}$, and the right panel to $D_L=200\,\mathrm{Mpc}$. Comparing the two panels, as $D_L$ increases, the peak value of the characteristic strain $h_{\mathrm{c}}(f)$ decreases, while the peak frequency remains essentially unchanged. Thus, $D_L$ mainly rescales the overall amplitude of the characteristic strain without changing the spectral shape. For $D_L=2\,\mathrm{Gpc}$ (left panel), the signal only approaches the BBO sensitivity curve and falls significantly below the curves of LISA, eLISA, TianQin, and DECIGO. For $D_L=200\,\mathrm{Mpc}$ (right panel), the signal lies above the sensitivity curves of BBO and DECIGO and is comparable to the LISA sensitivity curve, while remaining below those of eLISA and TianQin in the millihertz range.

\begin{figure*}
\includegraphics[width=1.025\textwidth]{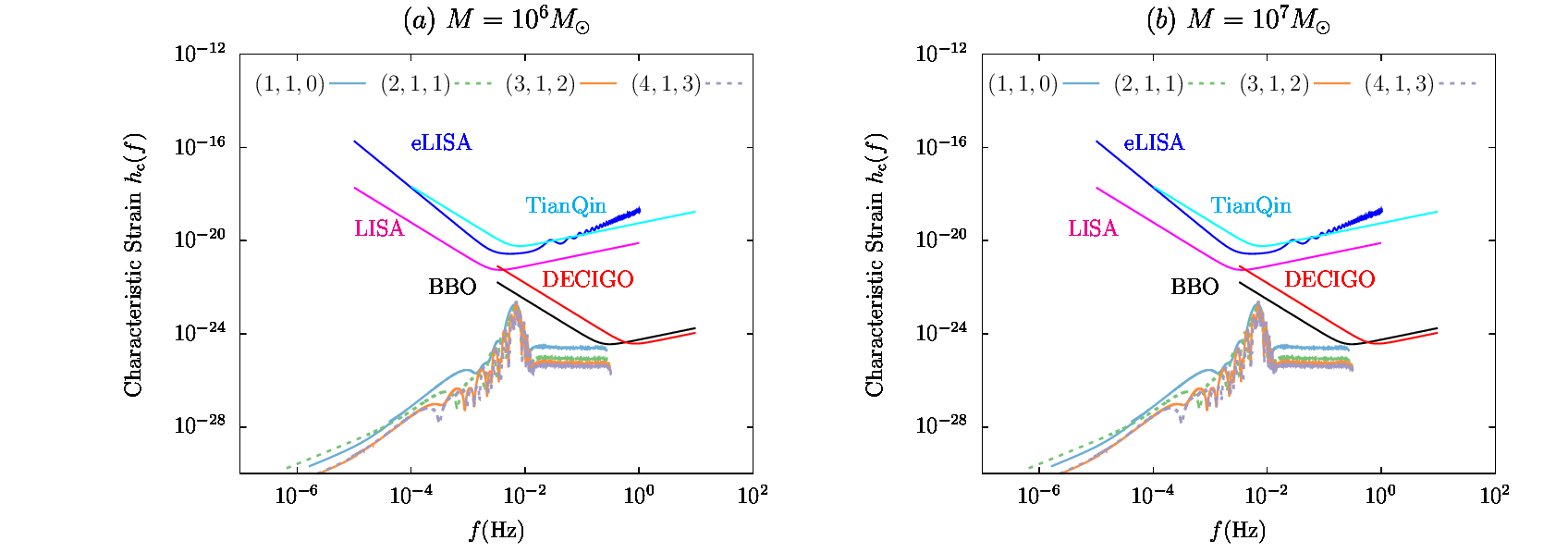}
\caption{Characteristic strain of gravitational waves from EMRIs on periodic orbits around Bardeen black holes, compared with the sensitivity curves of LISA (magenta), eLISA (blue), TianQin (cyan), BBO (black), and DECIGO (red). The left and right panels correspond to $M=10^6M_{\odot}$ and $M=10^7M_{\odot}$, respectively. The remaining parameters ($m$, $D_L$, $\iota$, $\zeta$) are fixed as in Fig.~\ref{fig:4}.
\label{fig:10}}
\end{figure*}

Figure~\ref{fig:10} presents the characteristic strain for fixed $m=10M_{\odot}$, $D_L=2\,\mathrm{Gpc}$, and $\iota=\zeta=\pi/4$, while varying the central black hole mass $M$. The left panel corresponds to $M=10^6M_{\odot}$, and the right panel to $M=10^7M_{\odot}$. The characteristic strain curves in the two panels are almost identical; the peak frequencies remain in the millihertz band around $10^{-2}\,\mathrm{Hz}$, and the $h_{\mathrm{c}}(f)$ values are of the same order. For $m\ll M$, the symmetric mass ratio satisfies $\eta\approx m/M$, so $h_{\mathrm{c}}(f) \propto m/D_L$ to leading order, independent of $M$. For both masses, the signals lie below the sensitivity curves of LISA, eLISA, TianQin, BBO, and DECIGO; therefore, none of these detectors is expected to detect the signal at this luminosity distance ($D_L=2\,\mathrm{Gpc}$).

\begin{figure*}
\includegraphics[width=1.025\textwidth]{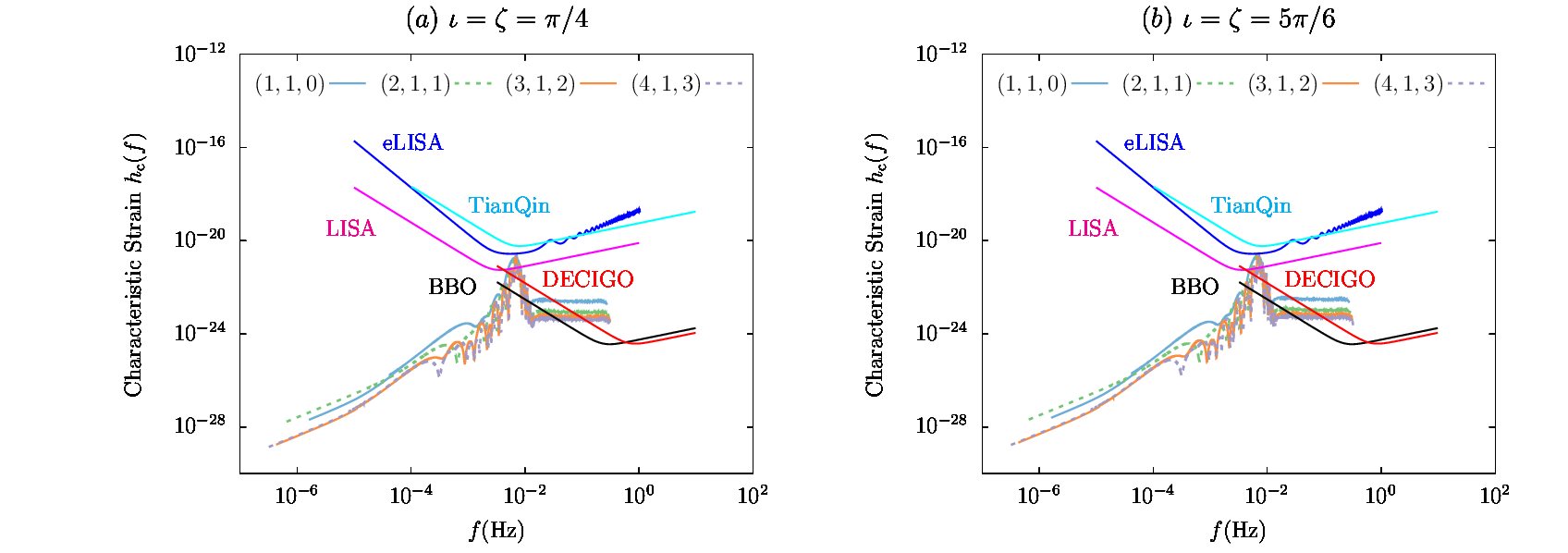}
\caption{Characteristic strain of gravitational waves from EMRIs on periodic orbits around Bardeen black holes, compared with the sensitivity curves of LISA (magenta), eLISA (blue), TianQin (cyan), BBO (black), and DECIGO (red). The left and right panels correspond to $\iota=\zeta=\pi/4$ and $\iota=\zeta=5\pi/6$, respectively. The remaining parameters ($m$, $M$, $D_L$) are fixed as in Fig.~\ref{fig:5}. 
\label{fig:11}}
\end{figure*}

Figure~\ref{fig:11} displays the characteristic strain for fixed $m=10M_{\odot}$, $M=10^6M_{\odot}$, and $D_L=200\,\mathrm{Mpc}$, while varying the inclination angle $\iota$ and the longitude of periastron $\zeta$. The left panel corresponds to $\iota=\zeta=\pi/4$, and the right panel to $\iota=\zeta=5\pi/6$. In both panels, the characteristic strain curves are almost identical: the peak frequency remains around $10^{-2}\,\mathrm{Hz}$ and the peak value remains on the order of $10^{-21}$. Only very small differences appear in the relative harmonic content and fine structure, much weaker than the changes induced by $D_L$. This is because $\iota$ enters only through the polarization amplitude coefficients $(1+\cos^2\iota)$ and $\cos\iota$, while $\zeta$ appears only a phase shift $2\zeta$. After combining the two polarizations into the characteristic strain, these changes mostly amount to a slight rescaling and phase rearrangement, without shifting the dominant frequency or changing the overall amplitude by orders of magnitude. Consequently, for both cases the signal lies above the BBO and DECIGO sensitivity curves and is comparable to the LISA sensitivity curve, but remains below those of eLISA and TianQin, so the overall detectability remains nearly unchanged.

These analyses show that the detectability of these GW signals depends primarily on the luminosity distance. For $m\ll M$, the dependence on the central black hole mass is weak, while $\iota$ and $\zeta$ produce only mild changes in the GW waveform. Space-based detectors such as LISA, eLISA, TianQin, BBO, and DECIGO are particularly sensitive to the low-frequency signals generated by these EMRIs, although the actual detectability varies with the source parameters. The corresponding comparisons are shown in Figs.~\ref{fig:9}, \ref{fig:10}, and \ref{fig:11}.

\section{Summary and Outlook}
\label{sec:5}

In this paper, we have presented a method to describe the geodesic motion and to characterize circular orbits in the $(L,E)$ parameter space for the Bardeen black hole spacetime. The method combines the extended periodic orbit classification with the analytical parametrization in terms of the orbital eccentricity $e$ and latus rectum $\lambda$, providing a systematic framework for generating periodic orbits and computing their GW signatures. While previous studies have mainly focused on the orbital classification and stability of bound geodesics in Bardeen black hole spacetimes, here we extend such analyses to the GW signatures of periodic orbits and assess their potential detectability with planned space-based detectors. The geodesic equations admit compact solutions in terms of elliptic integrals of the first kind. By parametrizing the motion with $e$ and $\lambda$, the energy and angular momentum for any periodic orbit specified by $(z,w,v)$ and $e$ can be obtained directly. This analytical approach enables us to efficiently generate the orbital configurations and subsequently compute their associated GW signatures. We note that our analysis is restricted to non-rotating Bardeen black holes and test particle orbits, neglecting the effects of black hole spin, radiation reaction, and astrophysical environments.

The main focus of this paper is the study of gravitational waves from EMRIs on periodic orbits around Bardeen black holes and the detectability of these waves. Our study reveals that the characteristic strain $h_{\mathrm{c}}(f)$ of GWs from these EMRIs depends strongly on the $(z,w,v)$ and the luminosity distance $D_{L}$. In the time domain, the luminosity distance has a significant effect on the waveform amplitude, while the central black hole mass, the inclination angle, and the longitude of periastron have only a negligible effect on the strain amplitude. In the frequency domain, the luminosity distance primarily rescales the overall amplitude of the spectrum, while the central black hole mass has only a weak effect on the spectrum. The inclination angle and the longitude of periastron mainly affect the relative harmonic content and spectral details. By comparing our numerical results with the sensitivity curves of planned or proposed space-based detectors (LISA, eLISA, TianQin, BBO, and DECIGO), we find that the low-frequency GW signals from these periodic orbits can be comparable to the sensitivity curves of several of these detectors, particularly LISA, BBO, and DECIGO, for sources at luminosity distance up to $200\,\mathrm{Mpc}$ and for Bardeen black hole masses around $10^6M_{\odot}$, while remaining below the eLISA and TianQin sensitivity curves in the millihertz range.

The main results can be summarized as follows: (i) We derived the analytical parametrization of periodic orbits in terms of the orbital eccentricity and latus rectum, enabling the construction of periodic orbits with the desired $(z,w,v;e)$ classification. (ii) We computed the time domain and frequency domain GW waveforms, showing that the characteristic strain is strongest at lower frequencies and varies significantly with the orbital parameters and source distance. (iii) We demonstrated that the characteristic strains of these EMRIs can be comparable to the sensitivity curves of LISA, BBO, and DECIGO, especially for sources at a luminosity distance of $200\,\mathrm{Mpc}$, while remaining below those of eLISA and TianQin, suggesting that these signals are potentially detectable by some of these detectors.

In summary, our findings indicate that periodic orbits around Bardeen black holes serve as excellent laboratories for probing gravity in the strong-field regime. The distinct GW signatures from these orbits could help probe deviations from the Kerr paradigm and constrain the properties of regular black hole spacetimes. In the future, we plan to extend this analysis to rotating Bardeen black holes, where the spin parameter will introduce additional complexity into the orbital dynamics and the resulting GW waveforms. A natural next step is to incorporate radiation reaction effects and perform a fully relativistic inspiral evolution, which would allow us to connect the periodic orbit waveforms to realistic EMRI signals. Furthermore, we intend to incorporate more realistic astrophysical environments, such as accretion disks, to better assess the detectability of these systems. It would also be interesting to perform Bayesian parameter estimation to quantify how well the magnetic monopole charge and other spacetime parameters can be constrained by future GW observations. The increasing precision of multi-messenger facilities such as the Event Horizon Telescope (EHT), Athena, and the Enhanced X-ray Timing and Polarimetry mission (eXTP) will make Bardeen black hole spacetimes a promising framework for probing regular black hole scenarios in the strong-field regime. Combining GW measurements with electromagnetic observations may further break parameter degeneracies and provide complementary tests of regular black hole scenarios. More broadly, our results highlight the potential of periodic orbit GW signals as a new window into the nature of compact objects and the strong-field regime of gravity, where deviations from the Kerr paradigm may be most likely to appear.

\begin{acknowledgements}
This work is supported by the National Natural Science Foundation of China, Grant Nos.11903022 and 12163003. 
\end{acknowledgements}

\bibliographystyle{spphys}
\bibliography{Refs20261012}

\end{document}